\documentclass[aps,manuscript,english]{revtex4}
\usepackage[T1]{fontenc}
\usepackage[latin1]{inputenc}
\usepackage{graphics}
\usepackage{times}
\makeatletter

\providecommand{\LyX}{L\kern-.1667em\lower.25em\hbox{Y}\kern-.125emX\@}

\usepackage[T1]{fontenc}
\usepackage[latin1]{inputenc}
\usepackage{graphics}

\makeatletter

\usepackage[T1]{fontenc}
\usepackage[latin1]{inputenc}
\usepackage{graphicx}

\makeatletter

\usepackage[T1]{fontenc}
\usepackage[latin1]{inputenc}
\usepackage{babel}
\usepackage{subfigure}



\makeatother
\makeatother
\makeatother

\begin{document}

\title{Macroscopic description of preheating}

\author{Esteban Calzetta}

\thanks{e-mail:calzetta@df.uba.ar }

\author{and Marc Thibeault }

\thanks{email:marc@df.uba.ar }

\affiliation{ Departamento de Física, UBA,Buenos Aires, Argentina}

\begin{abstract}
We present a macroscopic model of the decay of a coherent classical scalar field
into statistical fluctuations through the process of parametric amplification.
We solve the field theory (henceforth, \char`\"{}microscopic\char`\"{}) model
to leading order in a Large N expansion, and show that the macroscopic model
gives satisfactory results for the evolution of the field, its conjugated momentum
and the energy momentum tensor of the fluctuations over many oscillations. The
macroscopic model is substantially simpler than the microscopic one, and can
be easily generalized to include quantum fluctuations. Although we assume here
an homogeneous situation, the model is fully covariant, and can be applied in
inhomogeneous cases as well. These features make this model a promising tool
in exploring the physics of preheating. 
\end{abstract}

\maketitle

\section*{Introduction}

The eras of pre and reheating after inflation (\cite{preheating}) are generally
regarded as \char`\"{}probably the most violent of putative phases in cosmic
history\char`\"{}(\cite{Bassett}). During them, the quantum (maybe effective)
degree of freedom describing the inflaton decays into quantum and statistical
fluctuations of both gravitational and matter fields. The main decay mechanism
is parametric amplification of the fluctuations by the coherent oscillations
of the inflaton.

A full description of these phenomenon therefore requires an understanding of
the quantum field theory of parametric amplification (\cite{D.Boyanovsky})
, placed on a curved background, and including the backreaction on the geometry
(\cite{Ramseys}). Although there has been substantial progress in later years
(\cite{paper0},\cite{paper1},\cite{paper2},\cite{paper2b},\cite{paper3},\cite{Son})
many open questions remain, such as the relevance of nonlinear effects (\cite{Finelli})
and the generation of super-Hubble perturbations (\cite{Ashfordi}).

Further progress is hampered by many reasons, among which we believe the most
important one is that we do not really know the correct microscopic theory of
inflation. For this reason, it is necessary to consider a large number of competing
scenarios, giving divergent results for such a complex phenomenon as reheating.

However, this variability is limited by two factors. First, for the purposes
of cosmology we do not really need a detailed description of the process. Typically
what is required is the overall evolution of the inflaton field on large scales,
the final temperature of the radiation field - which dominates the specific
heat of the Universe after inflation - and an understanding of the time scales
involved. We may entertain the notion that different models may agree on these
very coarse-grained observables, while diverging under more sophisticated probes.

Second, while diverse, the models to be considered are not arbitrary. They must
be consistent with such general principles as causality and the Second Law of
thermodynamics. We know from relativistic field theory that even these very
general principles put nontrivial constraints on macroscopic behavior (\cite{fof},\cite{fof2}).

These observations suggest that it may be possible to investigate the model
independent, or at least robust, features of pre and reheating by replacing
the full microscopic models by simpler, macroscopic models answering to the
same constraints. The macroscopic model must be consistent with causality and
the Second Law, fully covariant, respect the conservation laws in the microscopic
model, and reproduce its equilibrium behavior. At the same time, the macroscopic
model should be a consistent hydrodynamical model by itself (\cite{Kraichnan}).
As we shall see in the following, these requirements alone essentially define
the macroscopic model. The remaining freedom concerns the values of the parameters
in the theory, which must be found from (numerical) experiment.

An indirect confirmation of the feasibility of this approach is the success
of a simple fluid model in reproducing some features of reheating (\cite{Grana}).
The requirement of full covariance must be stressed, since assuming a Friedmann
- Robertson - Walker background or low order perturbations thereof is inappropriate
to analyze the evolution of super-Hubble fluctuations.

In this paper we shall test these ideas by studying the decay of a coherent
self-interacting field into statistical fluctuations, both from the microscopic
equations of motion, and through a macroscopic model built according to the
principles we have set up.

In order to obtain a compelling result, we have chosen a well known problem,
namely the decay of an \( O(N) \) invariant scalar field, studied to leading
order in a Large N approximation (\cite{CHKMPA94},\cite{CKMP95},\cite{BVHS99},\cite{LMR02},\cite{HarHor81},\cite{MP89},\cite{RY00}\cite{AABBS02}).
In a certain sense, the leading order approximation is harder to study within
our approach than higher approximations would be. The leading order system is
Hamiltonian (see below) and therefore it is not associated to entropy production.
If integrated long enough, it shows revivals. The macroscopic model, being a
very coarse - grained version of the microscopic model, has a nontrivial entropy
production rate, and no revivals. Therefore any agreement between them can only
hold for times which are short with respect to the recurrence time. Higher order
approximations thermalize (\cite{BS03}) and it is possible to introduce a growing
entropy already at the microscopic level (\cite{Hu}), making them behave more
closely as the macroscopic model does.

Besides restricting ourselves to the leading order theory, we make further simplifications.
We shall assume homogeneous initial conditions in Minkowsky space, neglecting
the geometrical aspects of reheating, and we shall consider only a classical
field. This later approximation is sufficient for the study of reheating, as
recently shown by more complete analysis (\cite{BS02}). We assume that at the
initial time the fluctuations are in thermal equilibrium, although not in equilibrium
with the scalar field. We disregard initial time singularities (\cite{Baacke}).
We have presented elsewhere the general construction of the macroscopic model
(\cite{ours2}). The difference between this paper and those is that here we
lay the stress on showing that in this way it is possible to reproduce (in a
much more economical way) the overall behavior of a prescribed microscopic model.
The macroscopic model is based on Geroch's \char`\"{}Dissipative type theories\char`\"{}
(DTT) (\cite{Geroch}).

This framework ensures covariance and causality. To determine a concrete model
it is necessary to specify one thermodynamic potential and the form of the entropy
production. The former is found by requiring that the energy momentum tensor
as derived from the macroscopic model reproduces the expectation value of the
microscopic energy momentum tensor at the initial time. The entropy production
is severely restricted by demanding that the macroscopic model reproduces the
equation of motion for the mean field and its momentum at the initial time,
and that a vanishing field is a stable fixed point of the macroscopic equations.
This leaves only one decay constant undetermined, which is found by fitting
to the numerical results.

As we shall show, the macroscopic model succeeds in reproducing the evolution
of the mean field and the equivalent fluctuation temperature on time scales
short compared to the recurrence time, but long enough that initial correlations
among fluctuations are washed out. This result suggests that this approach could
be useful in exploring fully covariant, nonlinear reheating scenarios.

The rest of the paper is organized as follows.

Next Section is devoted to the microscopic model and its solution. We gather
here some known results which are necessary for comparison to the macroscopic
model later on. We derive the effective action from first principles and take
the limit \( \hbar \rightarrow 0 \) to obtain the energy-momentum tensor of
the field and it\'{s} thermal fluctuations which is our quantity of prime interest.

Section III presents the macroscopic model. After a short review of relativistic
thermodynamics (\cite{israel88},\cite{Calzetta}) we present the basics of
the DTT approach. We then discuss how to incorporate within these formalism
the fluctuation field, the mean field, and their interactions. We proceed by
identifying the thermodynamic potential from comparison of the respective energy
momentum tensors at the initial time. We show that the model is causal.

In Section IV we complete the derivation of the macroscopic model by analyzing
the entropy production rate. We present our results in the last section, where
both models are compared in detail.

A brief Appendix on DTT theory is attached. Appendix B contains some specific
calculations.

\section{The microscopic model }

\subsection{Equations of motions}

The starting point for our microscopic model is the \( O(N) \) invariant action
(\cite{D.Boyanovsky},\cite{CHKMPA94})

\begin{equation}
\label{old1}
S=N\int \left\{ -\frac{1}{2}\partial _{\mu }\varphi ^{i}\partial ^{\mu }\varphi ^{i}-\frac{\lambda }{2}\left( \frac{1}{2}\varphi ^{i}\varphi ^{i}+\frac{m^{2}}{\lambda }\right) ^{2}\right\} d^{4}x
\end{equation}

which represent a autointeracting field under the potential 
\begin{equation}
\label{pot}
V(\phi ^{i}\phi ^{i})=\frac{1}{2}m^{2}\phi ^{i}\phi ^{i}+\frac{1}{8}\lambda \left( \phi ^{i}\phi ^{i}\right) ^{2}
\end{equation}

the \( m^{4}/2\lambda  \) term is for convenience only, producing no change
in the equation of motion. It is customary to define a new field by adding a
constraint 
\begin{equation}
\label{sup1}
\frac{\left[ \chi -\lambda \left( \varphi ^{i}\varphi ^{i}+m^{2}/\lambda \right) \right] ^{2}}{2\lambda }
\end{equation}

which leads to a new action 
\begin{equation}
\label{sup2}
S_{2}=N\int \left\{ -\frac{1}{2}\partial _{\mu }\varphi ^{i}\partial ^{\mu }\varphi ^{i}+\frac{\chi ^{2}}{2\lambda }-\frac{1}{2}\chi \varphi ^{i}\varphi ^{i}-\frac{m^{2}}{\lambda }\chi \right\} d^{4}x
\end{equation}

Representing the expectation value of the field with respect to the initial
state of the theory by 
\begin{equation}
\label{sup3}
<\varphi >=\phi ,\qquad <\chi >=K
\end{equation}

we can write the fields as the sum of mean fields and fluctuations 
\begin{equation}
\label{old2}
\varphi ^{i}=\phi ^{i}+\psi ^{i}
\end{equation}

\begin{equation}
\label{sup4}
\chi =K+\kappa 
\end{equation}

Keeping only the leading term in the large \( N \) approximation we can write
the effective action 
\begin{eqnarray}
\Gamma [\phi ^{i},K] & = & S_{2}[\phi ^{i},K]+i\frac{N\hbar }{2}Tr\ln \left( \frac{i}{2\hbar }\right) \left( \partial _{\mu }\partial ^{\mu }-K\right) \nonumber \\
 & = & S_{2}[\phi ^{i},K]+i\frac{N\hbar }{2}Tr\ln \left( -\frac{1}{2}\right) G_{F}^{-1}(x-z)\label{old3} 
\end{eqnarray}

Variation of the effective action gives the equation for the classical fields
\( \phi ^{i} \)
\begin{equation}
\label{old4}
-\partial _{\mu }\partial ^{\mu }\phi ^{i}+K\phi ^{i}=0
\end{equation}

and for the \( K \) field, 
\begin{equation}
\label{old5}
K=m^{2}+\frac{\lambda }{2}\phi ^{i}\phi ^{i}+\frac{\lambda }{2}G_{F}(x,x)
\end{equation}

It is helpful at this point to rotate in the internal space so that \( \phi ^{i}=0 \)
for \( i\in [i,N-1] \), and \( \phi ^{N}\equiv \phi =\sqrt{\phi ^{i}\phi ^{i}} \).
Albeit it is true that the longitudinal and transverse fluctuations are different,
to first order in \( 1/N \) the Feynman propagator for the fluctuations is
given by 
\begin{equation}
\label{old6}
G_{F}(x,x^{\prime })\delta ^{ij}=<T\left( \psi ^{i}(x)\psi ^{j}(x^{\prime })\right) >
\end{equation}

which is solution of the following equation: 
\begin{equation}
\label{sup5}
\left( -\partial _{\mu }\partial ^{\mu }+K\right) G_{F}\left( x,x^{\prime }\right) =-i\hbar \delta \left( x-x^{\prime }\right) 
\end{equation}

If we assume a homogeneous initial state, it is convenient to introduce the
Fourier expansion 
\begin{equation}
\label{old7}
\psi ^{i}(x)=\int \frac{d^{3}k}{\left( 2\pi \right) ^{3}}\exp \left( i\vec{k}\cdot \vec{x}\right) \sqrt{\frac{\hbar }{2\omega _{k}(0)}}\left[ U_{k}(t)a_{\vec{k}}^{i}+U^{*}_{k}(t)a^{i}_{-\vec{k}}\right] 
\end{equation}

The normalization for the modes is

\begin{equation}
\label{old8}
W\left[ U_{k}^{*}(t),U_{k}(t)\right] =-i\hbar 
\end{equation}

\( W\left[ f,,g\right] =f\dot{g}-\dot{f}g \) being the Wronskian. In the homogeneous
case, 
\begin{equation}
\label{old8a}
\frac{d^{2}\phi }{dt^{2}}+K\phi =0
\end{equation}

with the corresponding equations of motion for the mode function \( U_{k}(t): \)
\begin{equation}
\label{old10}
\frac{d^{2}U_{k}(t)}{dt^{2}}+\omega _{k}^{2}(t)U_{k}(t)=0\quad ;\quad \omega _{k}^{2}(t)=|\vec{k}|^{2}+K(t)
\end{equation}

Substituting (\ref{old7}) en (\ref{old6}) gives 
\begin{equation}
\label{sup6}
G_{F}(x,x^{\prime })=G_{V}(x,x^{\prime })+G_{T}(x,x^{\prime })
\end{equation}

where we recognize a temperature-independent (or vacuum) part 
\begin{equation}
\label{sup7}
G_{V}(x,x^{\prime })=\int \frac{d^{3}k}{(2\pi )^{3}}\frac{\hbar }{2\omega _{k}(0)}\exp \left[ i\vec{k}\cdot (\vec{x}-\vec{x}^{\prime })\right] \left\{ U_{k}(t)U^{*}_{k}(t^{\prime })\Theta (t-t^{\prime })+U_{k}(t^{\prime })U^{*}_{k}(t)\Theta (t^{\prime }-t)\right\} 
\end{equation}

and a temperature -dependent part 
\begin{eqnarray}
G_{T}(x,x^{\prime }) & = & \int \frac{d^{3}k}{(2\pi )^{3}}\frac{\hbar }{2\omega _{k}(0)}\left\{ \exp \left[ i\vec{k}\cdot (\vec{x}+\vec{x}^{\prime })\right] \left[ U_{k}(t)U_{k}(t^{\prime })g_{k}+U^{*}_{k}(t)U^{*}_{k}(t^{\prime })g_{k}^{*}\right] \right. \\
 &  & \left. \exp \left[ i\vec{k}\cdot (\vec{x}-\vec{x}^{\prime })\right] \left[ U_{k}(t)U^{*}_{k}(t^{\prime })+U^{*}_{k}(t)U_{k}(t^{\prime })\right] n_{k}\right\} \label{sup8} 
\end{eqnarray}

where \( n_{k} \) and \( g_{k} \) represent the initial statistical mixture:
\begin{equation}
\label{sup9}
\left\langle a_{\vec{k}}^{i}\: a_{\vec{k}^{\prime }}^{j}\right\rangle =(2\pi )^{3}g_{k}\delta ^{3}(\vec{k}-\vec{k}^{\prime })\delta ^{ij}
\end{equation}

\begin{equation}
\label{sup10}
\left\langle a_{\vec{k}}^{i\dagger }\: a_{\vec{k}^{\prime }}^{j}\right\rangle =(2\pi )^{3}n_{k}\delta ^{3}(\vec{k}-\vec{k}^{\prime })\delta ^{ij}
\end{equation}

In the coincidence limit, 
\begin{equation}
\label{GV}
G_{V}(x,x)=\int \frac{d^{3}k}{(2\pi )^{3}}\frac{\hbar }{2\omega _{k}(0)}\left| U_{k}(t)\right| ^{2}
\end{equation}

and 
\begin{equation}
\label{GT}
G_{T}(x,x)=\int \frac{d^{3}k}{(2\pi )^{3}}\frac{\hbar }{2\omega _{k}(0)}\left\{ 2\exp \left( 2i\vec{k}\cdot \vec{x}\right) Re\left[ U_{k}(t)U_{k}(t)g_{k}\right] +2\left| U_{k}(t)\right| ^{2}n_{k}\right\} 
\end{equation}

We will assume a thermal bath of particles initially 
\begin{eqnarray}
n_{k} & = & \frac{1}{\exp \left( \frac{\hbar \omega _{k}(0)}{k_{B}T_{f}}\right) -1}\label{nk} \\
g_{k} & = & 0
\end{eqnarray}

with 
\begin{equation}
\label{Tf}
T=T_{f}
\end{equation}
 the initial temperature of the bath. The lowercase-index \( f \) is explicitly
used to prevent confusion later on and indicate the temperature of the initial
bath of the fluctuations.

\subsubsection{Classical Thermal fluctuations, energy-momentum tensor and Hamiltonian formulation }

A very successful strategy to deal with reheating in the strongly nonlinear
regime has been to describe the inflaton field as purely classical \cite{Tkachev00};
the rationale behind this approach is that there is a rapid transition to semiclassical
behavior during inflation \cite{Starobinsky95} and thus, if one succeeds in
finding a semiclassical description of fluctuations produced by the inflaton
decay, the decay itself can be described via classical equations of motion \cite{Tkachev96}.
In this paper we will focus on the particular case of classical thermal fluctuations.
This is a case that is easier to implement than the quantum one while allowing
us to confront the basic issues. We then take the limit \( \hbar \rightarrow 0 \)
in eq.(\ref{GV}) and (\ref{GT}) to obtain 
\begin{equation}
\label{sup11}
\lim _{\hbar \rightarrow 0}G_{V}(x,x)=\lim _{\hbar \rightarrow 0}\int \frac{d^{3}k}{(2\pi )^{3}}\frac{\hbar }{2\omega _{k}(0)}\left| U_{k}(t)\right| ^{2}=0
\end{equation}

\begin{eqnarray}
\lim _{\hbar \rightarrow 0}G_{T}(x,x) & = & \lim _{\hbar \rightarrow 0}\int \frac{d^{3}k}{(2\pi )^{3}}\frac{1}{2\omega _{k}(0)}\frac{\hbar }{\exp \left( \frac{\hbar \omega _{k}(0)}{k_{B}T_{f}}\right) -1}2\left| U_{k}(t)\right| ^{2}\label{sup12} \\
 & = & \int \frac{d^{3}k}{(2\pi )^{3}}\frac{2k_{B}T_{f}}{2\omega ^{2}_{k}(0)}\left| U_{k}(t)\right| ^{2}\label{sup13} 
\end{eqnarray}

Thus, in this classical limit, we find 
\begin{equation}
\label{old11}
K=m^{2}+\frac{1}{2}\lambda \phi ^{2}+\lambda k_{B}T_{f}\int \frac{d^{3}k}{(2\pi )^{3}}\frac{1}{2\omega ^{2}_{k}(0)}\left| U_{k}(t)\right| ^{2}
\end{equation}

The energy-momentum tensor can be computed using (\ref{old3}) and 
\begin{equation}
\label{sup14}
\Gamma ^{\mu \nu }=\frac{2}{\sqrt{-g}}\frac{\delta \Gamma }{\delta g_{\mu \nu }}
\end{equation}

specializing thereafter to Minkowski or directly by taking the expectation value
of the classical energy-momentum tensor. Either way, we find (writing only the
nontrivial components) 
\begin{eqnarray}
\left\langle T^{00}\right\rangle  & = & \frac{1}{2}\dot{\phi }^{2}+k_{B}T_{f}\int \frac{d^{3}k}{(2\pi )^{3}}\frac{1}{2\omega _{k}(0)}\left[ \dot{U}_{k}(t)\dot{U}^{*}_{k}(t)+|\vec{k}|^{2}\left| U_{k}(t)\right| ^{2}\right] \\
 &  & +\frac{1}{2\lambda }\left( K-m^{2}\right) \left( K+m^{2}\right) +\frac{m^{4}}{2\lambda }\label{sup15} 
\end{eqnarray}

and (no sum over \( i \) ) 
\begin{eqnarray}
\left\langle T^{ii}\right\rangle  & = & \frac{1}{2}\dot{\phi }^{2}+k_{B}T_{f}\int \frac{d^{3}k}{(2\pi )^{3}}\frac{1}{2\omega _{k}(0)}\left[ \dot{U}_{k}(t)\dot{U}^{*}_{k}(t)-\frac{1}{3}|\vec{k}|^{2}\left| U_{k}(t)\right| ^{2}\right] \\
 &  & -\frac{1}{2\lambda }\left( K-m^{2}\right) \left( K+m^{2}\right) -\frac{m^{4}}{2\lambda }\label{sup16} 
\end{eqnarray}

These expressions are the total energy density and pressure for our system.
To integrate numerically our equations we need a finite number of variables.
Since the integrands are manifestly isotropic, we perform first the integration
over the angular variables. In problems with spherical symmetry like this one,
this procedure leads to a better approximation of the integral than approximating
the 3-D integral as a triple sum over Cartesian coordinates. The remaining integral
is written as a finite sum: 
\begin{eqnarray}
\left\langle T^{00}\right\rangle  & = & \frac{1}{2}\dot{\phi }^{2}+k_{B}T_{f}\sum _{k=1}^{N_{k}}\Delta k\alpha _{(\Delta k\cdot k)}\left[ \dot{U}_{\Delta k\cdot k}(t)\dot{U}^{*}_{\Delta k\cdot k}(t)+|\vec{k}\Delta k|^{2}\left| U_{\Delta k\cdot k}(t)\right| ^{2}\right] \\
 &  & +\frac{1}{2\lambda }\left( K-m^{2}\right) \left( K+m^{2}\right) +\frac{m^{4}}{2\lambda }\label{old12} 
\end{eqnarray}

\begin{eqnarray}
\left\langle T^{ii}\right\rangle  & = & \frac{1}{2}\dot{\phi }^{2}+k_{B}T_{f}\sum _{k=1}^{N_{k}}\Delta k\alpha _{(\Delta k\cdot k)}\left[ \dot{U}_{\Delta k\cdot k}(t)\dot{U}^{*}_{\Delta k\cdot k}(t)-\frac{1}{3}|\vec{k}\Delta k|^{2}\left| U_{\Delta k\cdot k}(t)\right| ^{2}\right] \\
 &  & -\frac{1}{2\lambda }\left( K-m^{2}\right) \left( K+m^{2}\right) -\frac{m^{4}}{2\lambda }\label{old13} 
\end{eqnarray}

where we defined 
\begin{equation}
\label{alpha}
\alpha _{k}\equiv \frac{|\vec{k}|^{2}}{4\pi ^{2}\omega ^{2}_{k}(0)}
\end{equation}

and we set a cutoff frequency 
\begin{equation}
\label{cutoff}
k_{max}\equiv \Lambda =N_{k}\Delta _{k}.
\end{equation}
 \( N_{k} \) is the total number of modes and \( \Delta _{k} \) is the spacing.
It is straightforward exercise to verify that we can now rewrite the whole system
as a Hamiltonian system (\( H \) has units of energy density). Writing the
complex modes \( U_{k}(t) \) as real and imaginary parts 
\begin{equation}
\label{sup17}
U_{k}(t)=U_{r}(t)+iU_{i}(t)
\end{equation}

and writing \( p\equiv \dot{\phi } \) the Hamiltonian is 
\begin{eqnarray}
H & = & \frac{1}{2}p^{2}+\frac{1}{2}K\phi ^{2}+\sum _{k=1}^{N_{k}}\left\{ \frac{1}{4k_{B}T_{f}\alpha _{(\Delta _{k}\cdot k)}\Delta k}\left[ \Pi _{r}^{2}+\Pi _{i}^{2}\right] \right. \\
 &  & \left. +k_{B}T_{f}\alpha _{(\Delta _{k}\cdot k)}\Delta _{k}\left( |\vec{k}|^{2}+K\right) \left[ U_{r}^{2}+U_{i}^{2}\right] \right\} -\frac{1}{2\lambda }\left( K-m^{2}\right) ^{2}\label{old14} 
\end{eqnarray}

The equations of motions for the fluctuations are 
\begin{equation}
\label{motion1}
-\dot{\Pi }_{r}=\frac{\partial H}{\partial U_{r}}=2k_{B}T_{f}\alpha _{(\Delta _{k}\cdot k)}\Delta _{k}\left( |\Delta _{k}\cdot \vec{k}|^{2}+K\right) U_{r}
\end{equation}

\begin{equation}
\label{motion2}
\dot{U}_{r}=\frac{\partial H}{\partial \Pi _{r}}=\frac{1}{2k_{B}T_{f}\alpha _{(\Delta _{k}\cdot k)}\Delta _{k}}\Pi _{r}
\end{equation}

and 
\begin{equation}
\label{motion3}
-\dot{\Pi }_{i}=\frac{\partial H}{\partial U_{i}}=2k_{B}T_{f}\alpha _{(\Delta _{k}\cdot k)}\Delta _{k}\left( |\Delta _{k}\cdot \vec{k}|^{2}+K\right) U_{i}
\end{equation}

\begin{equation}
\label{motion4}
\dot{U}_{i}=\frac{\partial H}{\partial \Pi _{i}}=\frac{1}{2k_{B}T_{f}\alpha _{(\Delta _{k}\cdot k)}\Delta _{k}}\Pi _{i}
\end{equation}

which also give the definitions of \( \Pi _{r} \) and \( \Pi _{i} \). The
equations for the scalar field are 
\begin{equation}
\label{motion5}
\dot{\phi }=p
\end{equation}

and 
\begin{equation}
\label{motion6}
\dot{p}=-K\phi 
\end{equation}

Initially, we have \( U_{r}(0)=\cos (\delta ), \) \( \dot{U}_{r}(0)=\omega _{k}(0)\sin (\delta ) \)
and \( U_{i}(0)=\sin (\delta ), \) \( \dot{U}_{i}(0)=-\omega _{k}(0)\cos (\delta ) \)
where \( \delta  \) is some random phase. \( K \) is a Lagrange multiplier
and its corresponding Hamiltonian equation gives the (discretized) gap equation
(\ref{old11}). At \( t=0 \) this read

\begin{eqnarray}
K(0) & = & m^{2}+\frac{\lambda }{2}\phi ^{2}(0)+\lambda k_{B}T_{f}\sum _{k=1}^{N_{k}}\Delta _{k}\alpha _{\Delta _{k}\cdot k}\\
 & = & m^{2}+\frac{\lambda }{2}\phi ^{2}(0)+\frac{\lambda k_{B}T_{f}\Delta _{k}}{4\pi ^{2}}\sum _{k=1}^{N_{k}}\frac{|\Delta _{k}\vec{k}|^{2}}{\left( |\Delta _{k}\vec{k}|^{2}+K(0)\right) }\label{old15} 
\end{eqnarray}

an equation to be solved numerically to extract \( K(0) \) as a function of
the initial conditions and parameters. Note that discretization and imposing
a cutoff take care of both ultraviolet divergences and initial time singularities.
No further renormalization will be needed.

\subsection{The energy-momentum tensor at t=0}

As we are going to see later, in order to complete the macroscopic model we
need to know the exact form of the interaction potential. It would be very useful
to have an analytical expression for the total energy density and pressure in
the microscopic model. Our aim here is then to find an expression that is analytical,
manageable and a good approximation of our exact microscopic model.

Our scheme is the following. If one considers a dissipative fluid as a mixture
of two simpler fluids, most of the information specific to each one will be
lost after the mixing. We can only rely on quantities which can be computed
using the total energy density or pressure. However, at the initial time, there
is some specific information about each fluid that one can assumed to be known
from the initial conditions. On one hand, we know how to describe a classical
autointeracting field as a fluid (\cite{ours2}). On the other we know how to
theoretically describe fluctuations (quantum or thermal) as a fluid also. We
can then derive the contribution coming from the interaction potential, at the
very least at the initial time. We will suppose that the functional dependence
on the temperature is the same at later times. Comparison between the dynamics
of both models will validate this hypothesis. Under these premises, the three
quantities of interest are:

\begin{eqnarray}
H(t) & = & \frac{p^{2}}{2}+k_{B}T_{f}\int \frac{k^{2}dk}{4\pi ^{2}\omega ^{2}_{k}(0)}\left\{ \left| \dot{U}_{k}(t)\right| ^{2}+k^{2}\left| U_{k}(t)\right| ^{2}\right\} \\
 &  & +\frac{1}{2\lambda }\left( K(t)-m^{2}\right) \left( K(t)+m^{2}\right) +\frac{m^{4}}{2\lambda }\label{sup18} \\
P(t) & = & \frac{p^{2}}{2}+k_{B}T_{f}\int \frac{dk}{4\pi ^{2}\omega ^{2}_{k}(0)}\left\{ \left| \dot{U}_{k}(t)\right| -\frac{k^{2}}{3}\left| U_{k}(t)\right| ^{2}\right\} \\
 &  & -\frac{1}{2\lambda }\left( K(t)-m^{2}\right) \left( K(t)+m^{2}\right) -\frac{m^{4}}{2\lambda }\label{sup19} \\
K(t) & = & m^{2}+\frac{1}{2}\lambda \phi ^{2}+\lambda k_{B}T_{f}\int dk\frac{k^{2}}{4\pi ^{2}\omega ^{2}_{k}(0)}\left| U_{k}(t)\right| ^{2}\label{sup20} 
\end{eqnarray}

Where \( H=<T^{00}> \) represents the total energy density of the microscopic
theory and \( P=<T^{ii}> \) is the total pressure. The actual temperature of
the fluctuations at later times should be obtained using the gap equation and
some suitable guess on the modes. Using the initial conditions \( \left| \dot{U}_{k}(0)\right| \equiv \sqrt{\dot{U}_{r}^{2}(0)+\dot{U}^{2}_{i}(0)}=\omega _{k}(0) \)
y \( \left| U_{k}(0)\right| \equiv \sqrt{U^{2}_{r}(0)+U_{i}^{2}(0)}=1 \) we
have 
\begin{eqnarray}
H_{0} & = & \frac{p^{2}}{2}+k_{B}T_{f}\int \frac{k^{2}dk}{4\pi ^{2}\omega ^{2}_{k}(0)}\left\{ \omega ^{2}_{k}(0)+k^{2}\right\} \\
 &  & +\frac{1}{2\lambda }\left( K(0)-m^{2}\right) \left( K(0)+m^{2}\right) +\frac{m^{4}}{2\lambda }\label{sup21} \\
P_{0} & = & \frac{p^{2}}{2}+k_{B}T_{f}\int \frac{k^{2}dk}{4\pi ^{2}\omega ^{2}_{k}(0)}\left\{ \omega ^{2}_{k}(0)-\frac{k^{2}}{3}\right\} \\
 &  & -\frac{1}{2\lambda }\left( K(0)-m^{2}\right) \left( K(0)+m^{2}\right) -\frac{m^{4}}{2\lambda }\label{sup22} \\
K(0,\phi _{0}) & = & m^{2}+\frac{1}{2}\lambda \phi _{0}^{2}+\lambda k_{B}T_{f}\int dk\frac{k^{2}}{4\pi ^{2}\omega ^{2}_{k}(0)}\label{sup23} 
\end{eqnarray}

where we write explicitly the functional dependence of \( K(t=0) \) with \( \phi _{0} \)
and we use the notation \( H(t=0)\equiv H_{0} \) and similarly with the pressure.
Let\'{ }s compute first \( K(0,\phi _{0}). \) We can rewrite it as 
\begin{eqnarray}
K(0) & = & m^{2}+\frac{1}{2}\lambda \phi _{0}^{2}+\lambda k_{B}T_{f}\int dk\frac{k^{2}+K(0)-K(0)}{4\pi ^{2}\left( k^{2}+K(0)\right) }\label{sup24} \\
 & = & m^{2}+\frac{1}{2}\lambda \phi _{0}^{2}+\lambda k_{B}T_{f}\int dk\frac{1}{4\pi ^{2}}-\lambda k_{B}T_{f}\int \frac{dkK(0)}{4\pi ^{2}\left( k^{2}+K(0)\right) }\label{sup25} 
\end{eqnarray}

That is 
\begin{equation}
\label{K(0)}
\left\{ 1+\lambda k_{B}T_{f}\int \frac{dk}{4\pi ^{2}\left( k^{2}+K(0)\right) }\right\} K(0)=m^{2}+\frac{1}{2}\lambda \phi _{0}^{2}+\frac{\lambda k_{B}T_{f}}{4\pi ^{2}}\int dk
\end{equation}

The last integral is ill-defined, the culprit being the model itself: it is
the ultraviolet divergence which has its root in the classical equipartition
theorem. In our model, the divergence is controlled by assuming a finite cut-off
\( \Lambda . \) Alternatively, the theory can be renormalized via renormalization
of the mass (\cite{Aarts}). Integrating we found 
\begin{equation}
\label{kphi}
\left\{ 1+\frac{\lambda k_{B}T_{f}}{4\pi ^{2}}\frac{1}{\sqrt{K(0,\phi _{0})}}\arctan \frac{\Lambda }{\sqrt{K(0,\phi _{0})}}\right\} K(0,\phi _{0})=m^{2}+\frac{1}{2}\lambda \phi _{0}^{2}+\frac{\lambda k_{B}T_{f}}{4\pi ^{2}}\Lambda 
\end{equation}

A numerical solution will yield \( K(0,\phi _{0}). \) Note that, in the limit
\( T\rightarrow 0 \) we have \( K(0,\phi _{0})=m^{2}+\frac{1}{2}\lambda \phi _{0}^{2} \).
Let\'{ }s now turn to the total initial energy density. We have 
\begin{eqnarray}
H_{0}(T_{f}) & = & \frac{p^{2}}{2}+\frac{k_{B}T_{f}}{6\pi ^{2}}\Lambda ^{3}+\frac{k_{B}T_{f}}{4\pi ^{2}}\left( -\Lambda K(0,\phi _{0})+\left( K(0,\phi _{0})\right) ^{3/2}\arctan \left( \frac{\Lambda }{\sqrt{K(0,\phi _{0})}}\right) \right) \nonumber \\
 &  & +\frac{K(0,\phi _{0})^{2}}{2\lambda }\label{T00phi} 
\end{eqnarray}

indicating explicitly the dependency on the temperature. A similar calculation
yields for the total initial pressure: 
\begin{eqnarray}
P_{0}(T_{f}) & = & \frac{p^{2}}{2}+\frac{k_{B}T_{f}}{18\pi ^{2}}\Lambda ^{3}-\frac{k_{B}T_{f}}{12\pi ^{2}}\left( -\Lambda K(0,\phi _{0})+\left( K(0,\phi _{0})\right) ^{3/2}\arctan \left( \frac{\Lambda }{\sqrt{K(0,\phi _{0})}}\right) \right) \nonumber \\
 &  & -\frac{K(0,\phi _{0})^{2}}{2\lambda }\label{Tiiphi} 
\end{eqnarray}

It is possible to simplify these two expression using eq. (\ref{kphi}): 
\begin{equation}
\label{H0sim}
H_{0}(T_{f})=\frac{1}{2}p^{2}+\frac{k_{B}T_{f}\Lambda ^{3}}{6\pi ^{2}}+\frac{1}{2}K(0,\phi _{0})\phi _{0}^{2}+\frac{m^{2}K(0,\phi _{0})}{\lambda }-\frac{K^{2}(0,\phi _{0})}{2\lambda }
\end{equation}

\begin{equation}
\label{P0sim}
P_{0}(T_{f})=\frac{1}{2}p^{2}+\frac{k_{B}T_{f}}{18\pi ^{2}}\Lambda ^{3}-\frac{1}{6}K(0,\phi _{0})\phi _{0}^{2}-\frac{m^{2}K(0,\phi _{0})}{3\lambda }-\frac{K^{2}(0,\phi _{0})}{6\lambda }
\end{equation}
 The energy density of the fluctuations \( \rho _{q}, \) is obtained by considering
the above results at \( p_{0}=\phi _{0}=0 \) that is 
\begin{eqnarray}
\rho _{f} & = & \frac{k_{B}T_{f}}{6\pi ^{2}}\Lambda ^{3}+\frac{m^{2}K(0,0)}{\lambda }-\frac{K^{2}(0,0)}{2\lambda }-\frac{m^{4}}{2\lambda }\\
 & = & \frac{k_{B}T_{f}}{6\pi ^{2}}\Lambda ^{3}-\frac{1}{2\lambda }\left( K(0,0)-m^{2}\right) ^{2}\label{rhof} 
\end{eqnarray}

\vspace{0.3cm}

and 
\begin{eqnarray}
p_{f} & = & \frac{k_{B}T_{f}}{18\pi ^{2}}\Lambda ^{3}-\frac{m^{2}K(0,0)}{3\lambda }-\frac{K^{2}(0,0)}{6\lambda }+\frac{m^{4}}{2\lambda }\\
 & = & \frac{k_{B}T_{f}}{18\pi ^{2}}\Lambda ^{3}-\frac{1}{6\lambda }\left( K(0,0)-m^{2}\right) \left( K(0,0)+3m^{2}\right) \label{pf} 
\end{eqnarray}

Note that \( \rho _{f} \) and \( p_{f} \) are only defined up to an additive
constant. We used that small indeterminacy in their definitions in order to
have a natural limit for \( \rho _{f} \) and \( p_{f} \) for \( T_{f}\rightarrow 0. \)
Also note that \( K(0,0) \) indicates the solution of eq. (\ref{kphi}) with
\( \phi _{0}=0 \). In the case \( T_{f}\rightarrow 0 \), we find 
\begin{eqnarray}
H_{0} & = & \frac{p_{0}^{2}}{2}+\frac{1}{2\lambda }\left( m^{2}+\frac{1}{2}\lambda \phi _{0}^{2}\right) ^{2}\\
 & = & \frac{p_{0}^{2}}{2}+V(\phi _{0})+\frac{m^{4}}{2\lambda }
\end{eqnarray}

using (\ref{pot}) and 
\begin{equation}
\label{rhoqfi}
\rho _{f}=0=p_{f}\qquad \qquad (T_{f}=0)
\end{equation}

\section{The macroscopic model}

\subsection{Klein-Gordon field and fluctuations in the DTT framework }

\subsubsection{Relativistic Thermodynamics and DTT}

In order to motivate our formalism, we will present a very brief summary of
relativistic thermodynamics (\cite{israel88}). Recall that the basic thermodynamic
relations are coded in the Euler equation, which gives the entropy 
\begin{equation}
\label{entropy}
S=\frac{1}{T}\left[ U+pV-\mu N\right] 
\end{equation}
 and the first law 
\begin{equation}
\label{first law}
dU=TdS-pdV+\mu dN
\end{equation}

which ensures conservation of energy. Taking differential of (\ref{entropy})
and subtracting (\ref{first law}) we obtain the Gibbs-Duhem relation 
\begin{equation}
\label{sup26}
Vdp=SdT+Nd\mu 
\end{equation}
 from which we deduce 
\begin{equation}
\label{sup27}
s\equiv \frac{S}{V}=\left. \frac{\partial p}{\partial T}\right| _{\mu }\qquad ;\qquad n\equiv \frac{N}{V}=\left. \frac{\partial p}{\partial \mu }\right| _{T}
\end{equation}
 We transform this into a covariant theory by adapting the following rule (\cite{israel88}):

\begin{enumerate}
\item Intensive quantities \( (T,p,\mu ) \) are associated to scalars, which represent
the value of the quantity at a given event, as measured by an observer at rest
with respect to the fluid. 
\item Extensive quantities \( (S,V,N) \) are associated to vector currents \( S^{a},u^{a},N^{a} \)
such that given a timelike hypersurface element \( d\Sigma _{a}=n_{a}d\Sigma , \)
then \( -X^{a}d\Sigma _{a} \) is the amount of quantity \( X \) within the
volume \( d\Sigma  \) as measured by an observer with velocity \( n^{a} \).
If further the quantity \( X \) is conserved, then \( X^{a}_{\: ;a}=0. \)
The quantity \( u^{a} \) associated to volume \( d\Sigma _{a} \) is the fluid
four-velocity, and therefore obeys the additional constraint \( u^{2}=-1 \). 
\item Energy and momentum are combined into a single extensive quantity and associated
to a tensor current \( T^{ab}. \) The energy current, properly speaking, is
\( U^{a}=-T^{ab}u_{b}. \)
\end{enumerate}
According to this rule, eq. (\ref{entropy}) translates into 
\begin{equation}
\label{sup28}
TS^{a}=-T^{ab}u_{b}+pu^{a}-\mu N^{a}
\end{equation}

which we rewrite as 
\begin{equation}
\label{firstb}
S^{a}=\chi ^{a}-\beta _{b}T^{ab}-\alpha N^{a}
\end{equation}

where we introduced the affinity \( \alpha =\mu /T \) , the thermodynamic potential
\( \chi ^{a}=p\beta ^{a} \), and the inverse temperature vector \( \beta ^{a}=T^{-1}u^{a}. \)
We now introduce the concept of a perfect fluid, that is a system whose energy-momentum
tensor takes the form \( T^{ab}=g^{ab}+u^{a}u^{b}(\rho +p), \) where \( \rho  \)
is the energy density as seen by an observer moving with the fluid. Usually
this is not sufficient to characterize completely the fluid, and another equation
appears in the form of a conserved current \( N^{a}_{\; ;a}=0, \) where \( N^{a}=nu^{a}, \)
\( n \) being the corresponding density as seen by a comoving observer.

Without friction, that is all there is to it. When dissipation enters, things
become far more complicated and subtle. One would like to obtain an equivalent
to the Navier-Stokes equations that describe adequately the fluid, even when
the speed is near that of light. The proposals of Eckart and Landau where found
to fail to preserve causality and have stability problems (\cite{fof},\cite{fof2}).
To cure that defect, Geroch presented (\cite{Geroch}) a set of equations of
hyperbolic character (instead of the parabolic form of Navier-Stokes) that enforces
causality. A brief review of this formalism, is presented in appendix 1.

In Geroch\'{ }s approach, \( T^{ab} \) and \( n^{a} \) are assumed to be
derivable from a generating function \( \chi ^{a} \) : \( T^{ab}=\frac{\partial \chi ^{a}}{\partial \xi _{b}},\qquad n^{a}=\frac{\partial \chi ^{a}}{\partial \zeta } \)

\( \xi  \) and \( \xi ^{a} \) representing now the dynamical degrees of freedom
of the theory. \( \chi ^{a} \) can be further simplified since, as a consequence
of the symmetry of \( T^{ab}, \) we have 
\begin{equation}
\label{old16}
\chi ^{a}=\frac{\partial \chi }{\partial \xi _{a}}
\end{equation}
 That is, all the fundamental tensors of the theory can be obtained from the
generating functional \( \chi . \) As an example, a perfect fluid is obtained
if \( \chi =\chi (\zeta ,\mu ), \) where \( \xi =\sqrt{-\xi ^{a}\xi _{a}} \).
Simple differentiation gives 
\begin{equation}
\label{sup29}
T_{p}^{ab}=-\frac{g^{ab}}{\xi }\frac{\partial \chi _{p}}{\partial \xi }+\frac{\xi ^{a}\xi ^{b}}{\xi ^{2}}\left[ -\frac{1}{\xi }\frac{\partial \chi _{p}}{\partial \xi }+\frac{\partial ^{2}\chi _{p}}{\partial \xi ^{2}}\right] 
\end{equation}
 allowing the identification 
\begin{equation}
\label{dttp}
p=-\frac{1}{\xi }\frac{\partial \chi }{\partial \xi }
\end{equation}

\begin{equation}
\label{ddtrho}
\rho =\frac{\partial ^{2}\chi }{\partial \xi ^{2}}
\end{equation}

\subsubsection{Thermal fluctuations as a fluid~ }

Thermal fluctuations could be seen as a particular case of the preceding section;
the only complication is how to implement the cutoff. In the case of the fluctuations,
there is no chemical potential, and we are left with \( \beta ^{a} \) as the
thermodinamical degree of freedom. However, the cut-off condition selects a
special frame. Let \( \Upsilon ^{a} \) be the 4-vector associated to this frame.
We may think of \( \Upsilon ^{a} \) as the 4-velocity of the observers in whose
rest-frame the cut-off is imposed. Our generating potential will take the form
\( \chi _{f}=\chi _{f}\left( \xi ,\upsilon \right)  \) where \( \xi =\sqrt{-\beta _{a}\beta ^{a}} \)
and \( \upsilon =-\beta _{a}\Upsilon ^{a} \). Then 
\begin{equation}
\label{sup30}
\chi ^{a}=-\frac{\beta ^{a}}{\xi }\frac{\partial \chi }{\partial \xi }-\Upsilon ^{a}\frac{\partial \chi }{\partial \upsilon }
\end{equation}

and 
\begin{eqnarray}
\frac{\partial \chi ^{a}}{\partial \beta _{b}} & = & g^{ab}\left( -\frac{1}{\xi }\frac{\partial \chi }{\partial \xi }\right) +\frac{\beta ^{a}\beta ^{b}}{\xi }\frac{\partial }{\partial \xi }\left( \frac{1}{\xi }\frac{\partial \chi }{\partial \xi }\right) \\
 &  & +\frac{1}{\xi }\left( \beta ^{a}\Upsilon ^{b}+\Upsilon ^{a}\beta ^{b}\right) \frac{\partial ^{2}\chi }{\partial \xi \partial \upsilon }+\Upsilon ^{a}\Upsilon ^{b}\frac{\partial ^{2}\chi }{\partial \upsilon ^{2}}\label{sup31} 
\end{eqnarray}

For our goal, it is sufficient to choose the following ansatz for \( \chi  \)
\begin{equation}
\label{sup32}
\chi \left( \xi ,\upsilon \right) =\chi _{1}\left( \xi \right) +\chi _{2}\left( \upsilon \right) 
\end{equation}

Then 
\begin{equation}
\label{sup33}
T^{ab}=g^{ab}\left( -\frac{1}{\xi }\frac{\partial \chi _{1}}{\partial \xi }\right) +\frac{\beta ^{a}\beta ^{b}}{\xi }\frac{\partial }{\partial \xi }\left( \frac{1}{\xi }\frac{\partial \chi _{1}}{\partial \xi }\right) +\Upsilon ^{a}\Upsilon ^{b}\frac{\partial ^{2}\chi _{2}}{\partial \upsilon ^{2}}
\end{equation}

Therefore 
\begin{equation}
\label{sup34}
p_{f}=-\frac{1}{\xi }\frac{\partial \chi _{1}}{\partial \xi }
\end{equation}

but 
\begin{equation}
\label{sup35}
\rho _{f}=\frac{\partial ^{2}\chi _{1}}{\partial \xi ^{2}}+\left( \Upsilon ^{0}\right) ^{2}\frac{\partial ^{2}\chi _{2}}{\partial \upsilon ^{2}}
\end{equation}

To solve explicitly, we identify \( \xi  \) with \( \beta  \). Then 
\begin{equation}
\label{sup36}
\chi _{1}=-\int \beta p_{f}d\beta 
\end{equation}
 and 
\begin{equation}
\label{sup37}
\rho _{f}+\frac{\partial \left( \beta p_{f}\right) }{\partial \beta }=\left( \Upsilon ^{0}\right) ^{2}\frac{\partial ^{2}\chi _{2}}{\partial \upsilon ^{2}}
\end{equation}

As an example, if we used only the leading term we have 
\begin{equation}
\label{sup38}
p_{f}=\frac{\Lambda ^{3}}{18\pi ^{2}\beta }\rightarrow \chi _{1}=-\frac{\Lambda ^{3}}{18\pi ^{2}}\beta 
\end{equation}
 and, using \( \upsilon =-\beta _{a}\Upsilon ^{a}=\beta \Upsilon ^{0} \), 
\begin{equation}
\label{sup39}
\frac{\Lambda ^{3}}{6\pi ^{2}\beta }=\frac{\upsilon }{\beta }\frac{\partial ^{2}\chi _{2}}{\partial \upsilon ^{2}}
\end{equation}

we obtain 
\begin{equation}
\label{sup40}
\chi _{2}=\frac{\Lambda ^{3}}{6\pi ^{2}}\left( \upsilon \ln \upsilon -\upsilon \right) 
\end{equation}

Equivalently, one can write \( \upsilon =\beta  \) to perform the computation,
replacing \( \beta \rightarrow \upsilon  \) in the expression for \( \chi _{2} \)
afterward.

\subsubsection{DTT formulation of the the Klein-Gordon theory}

Casting an autointeracting scalar field in the framework of DTT is akin to the
description of Landau of a superfluid as a mixture of a normal fluid and a superfluid.
The super fluid has no entropy by itself and therefore cannot have a temperature
defined for it. The scalar field, as a coherent solution of its classical equation
of motion, will have zero entropy also and one thus expect difficulties in trying
to describe it as a classical fluid. However, one can define a larger theory
via the following equations 
\begin{equation}
\label{larger1}
T^{ab}_{\: ;b}=0
\end{equation}

and 
\begin{equation}
\label{larger2}
j^{a}_{\: ;a}=R
\end{equation}

adopting the following constitutive relation for the energy-momentum tensor
\begin{equation}
\label{larger3}
T^{ab}=j^{a}j^{b}-g^{ab}\left( \frac{1}{2}j_{c}j^{c}+T[R]\right) 
\end{equation}

The scalar field is introduced by writing the functional relationship between
\( R \) and \( T \) parametrically as \( R=V^{\prime }\left( \phi \right)  \)
and \( T=V(\phi ) \). If one forces the constraint 
\begin{equation}
\label{sup41}
j_{b,c}-j_{c,b}=0
\end{equation}

then \( j^{a}=\phi ^{,a} \) and we fall back on the usual Klein-Gordon theory.
One can introduce new variables \( \xi  \) and \( \beta ^{a}_{(c)} \) and
a generating functional \( \chi ^{a}=\beta ^{a}_{(c)}p. \) In (\cite{ours2})
we show that, the Klein-Gordon theory can be reobtained if 
\begin{equation}
\label{sup42}
\chi _{c}=-\frac{1}{2}\xi ^{2}\ln \beta _{(c)}+\frac{1}{2}V\left( \phi \right) \beta _{(c)}^{2}
\end{equation}
 with the current given by \( j^{a}=\frac{\xi }{\beta ^{2}_{(c)}}\beta ^{a}_{(c)}, \)
and 
\begin{equation}
\label{constitutive}
\xi =-\beta _{(c)}^{a}\phi _{,a}.
\end{equation}
 The Klein-Gordon equation is given by \( j^{a}_{\; ;a}=V^{\prime }(\phi ), \)
\( V(\phi ) \) being the potential. Direct computation yields \( S^{a}=0 \)
and \( S^{a}_{\: ;a}=0 \) as it should be. Note that in the \'{ }rest\'{ }
frame where \( \beta ^{a}_{(c)}=\left( \beta _{(c)},\vec{0}\right)  \) we see
that the canonical momentum \( p\equiv \dot{\phi } \) is given by \( p=-\xi /\beta _{(c)} \)
. In Klein-Gordon theory, only this ratio has meaning. In order to break this
indeterminacy, we must look at a larger framework where the Klein-Gordon field
interacts with another fluid. In this larger context, eq.(\ref{constitutive})
will be taken as the definition of the scalar field \( \xi . \) The theory
describing the scalar field and a perfect fluid (f-fluid) together can be obtained
enlarging the generating potential by the addition of an interaction functional
\( \Xi  \), as we explain in the following section.

\subsection{The mixture of the two perfect fluids}

After coupling the scalar field and fluctuations, the energy-momentum tensor
of field and fluid will not be individually conserved and furthermore the Klein-Gordon
equation will also deviate from its original form. The total set of equations
governing our theory can be written 
\begin{eqnarray}
j^{a}_{\; ;a} & = & V^{\prime }(\phi )+\Delta \label{nueva} \\
T^{ab}_{(c);b} & = & I^{a}\label{nueva2} \\
T^{ab}_{(f);b} & = & -I^{a}\label{nueva3} \\
\beta ^{a}_{(c)}\phi _{,a} & = & -\xi \label{old17} 
\end{eqnarray}

where a subindex \( c\: (f) \) indicates a quantity belonging to the scalar
field (thermal fluctuations). \( j^{a} \) and \( T_{(c)}^{ab} \) are the current
and energy-momentum tensor for the scalar field. In this paper, we will work
under the assumption of homogeneity, which will reduce the number of equations
for the fluid to four. The total system, consisting in the union of the two
(perfect) fluids, differs in two ways from the initial q and c-fluids. First,
the generating functional of the theory is generated not only by the sum of
the generating functional of each fluid. There is also a third potential to
include the interaction between the two fluids 
\begin{equation}
\label{sup43}
\chi ^{a}=\chi ^{a}_{(c)}+\chi ^{a}_{(f)}+\Xi ^{a}
\end{equation}
 Thus

\begin{equation}
\label{sup44}
T^{ab}_{(c)}=\frac{\partial (\chi _{(c)}^{a}+\Xi ^{a})}{\partial \beta _{(c)b}}
\end{equation}

\begin{equation}
\label{sup45}
T^{ab}_{(f)}=\frac{\partial (\chi _{(f)}^{a}+\Xi ^{a})}{\partial \beta _{(f)b}}
\end{equation}

Moreover, the mixing will also generate entropy. The entropy production can
be computed using the equations of motion: 
\begin{equation}
\label{sup46}
\nabla _{a}S^{a}=-\xi \Delta -B_{b}I^{b}
\end{equation}

The pressure is by definition given by 
\begin{equation}
\label{sup47}
\Pi \equiv T^{ii}
\end{equation}

In the homogeneous case, the only nontrivial independent equation for the energy-momentum
tensor is the \( a=b=0 \) one. It is convenient to work with 
\begin{equation}
\label{old18}
T^{ab}_{+}=T^{ab}_{(c)}+T^{ab}_{(f)}=\frac{\partial \chi ^{a}}{\partial \beta _{b}}
\end{equation}

\begin{equation}
\label{old19}
T^{ab}_{-}=T^{ab}_{(c)}-T^{ab}_{(f)}=2\frac{\partial \chi ^{a}}{\partial B_{b}}
\end{equation}
 where the new variables \( B^{a}=\beta ^{a}_{(c)}-\beta ^{a}_{(f)} \) and
\( 2\beta =\beta ^{a}_{(c)}+\beta ^{a}_{(f)}. \) To compute (\ref{old18})
and (\ref{old19})we need to propose some specific form for the interaction
functional \( \Xi . \) We use as model the following interaction potential
\cite{ours2}: 
\begin{equation}
\label{intpot}
\Xi =F(u)v+G(u)w^{2}
\end{equation}
 which can be understood as a lowest order Taylor development in \( B^{2} \)
assuming moreover that it is depending only on the temperatures. We have introduced
the following variables: 
\begin{eqnarray}
u & = & -\beta _{a}\beta ^{a}=\beta ^{2}\label{sup48} \\
v & = & -B_{a}B^{a}=B^{2}\label{sup49} \\
w & = & -B_{a}\beta ^{a}=\beta B\label{sup50} 
\end{eqnarray}

which represent the scalars that can be made from the mean inverse temperature
\( \beta  \) and the inverse temperature difference \( B \). Starting from
(\ref{intpot}) we computed elsewhere (\cite{ours2}\textbf{)}
\begin{equation}
\label{sup51}
\Xi ^{a}=-2\beta ^{a}\left( \frac{dF}{du}v+\frac{dG}{du}w^{2}\right) -2B^{a}Gw
\end{equation}

\begin{eqnarray}
\frac{\partial \Xi ^{a}}{\partial \beta _{b}} & = & 4\beta ^{a}\beta ^{b}\left( \frac{d^{2}F}{du^{2}}v+\frac{d^{2}G}{du^{2}}w^{2}\right) -2g^{ab}\left( \frac{dF}{du}v+\frac{dG}{du}w^{2}\right) \\
 &  & +4\left( \beta ^{a}B^{b}+B^{a}\beta ^{b}\right) \frac{dG}{du}w+2B^{a}B^{b}G\label{sup52} 
\end{eqnarray}
 It is straightforward from there to calculate 
\begin{equation}
\label{sup53}
\frac{\partial \Xi ^{0}}{\partial \beta _{0}}=\left\{ 4\beta ^{2}\frac{d^{2}F}{du^{2}}+2\frac{dF}{du}+4\beta ^{4}\frac{d^{2}G}{du^{2}}+10\beta ^{2}\frac{dG}{du}+2G\right\} B^{2}
\end{equation}

\begin{equation}
\label{sup54}
\frac{\partial \Xi ^{i}}{\partial \beta _{i}}=-2\left( \frac{dF}{du}+\beta ^{2}\frac{dG}{du}\right) B^{2}
\end{equation}

It is easier to work directly with the \( \beta - \)derivative. We have 
\begin{equation}
\label{Xprime}
\frac{dF}{du}=\frac{d\sqrt{u}}{du}\frac{dF}{d\beta }=\frac{1}{2\beta }\frac{dF}{d\beta }
\end{equation}

\begin{equation}
\label{Xprime2}
\frac{d^{2}F}{du^{2}}=-\frac{1}{4\beta ^{3}}\frac{dF}{d\beta }+\frac{1}{4\beta ^{2}}\frac{d^{2}F}{d\beta ^{2}}
\end{equation}

Therefore 
\begin{equation}
\label{sup55}
\frac{\partial \Xi ^{0}}{\partial \beta _{0}}=\left\{ \frac{d^{2}F}{d\beta ^{2}}+\beta ^{2}\frac{d^{2}G}{d\beta ^{2}}+4\beta \frac{dG}{d\beta }+2G\right\} B^{2}
\end{equation}

and 
\begin{equation}
\label{sup56}
\frac{\partial \Xi ^{i}}{\partial \beta _{i}}=-\left( \frac{1}{\beta }\frac{dF}{d\beta }+\beta \frac{dG}{d\beta }\right) B^{2}
\end{equation}

Instead of working with \( F \) and \( G \), we will define the dimensionless
\( f \) and \( g \) as follows 
\begin{equation}
\label{Ff}
F=\beta ^{-4}f
\end{equation}

\begin{equation}
\label{Gg}
G=\beta ^{-6}g
\end{equation}

That is 
\begin{equation}
\label{chibeta0}
\frac{\partial \Xi ^{0}}{\partial \beta _{0}}=\frac{d^{2}}{d\beta ^{2}}\left( \frac{f+g}{\beta ^{4}}\right) B^{2}
\end{equation}

Also 
\begin{equation}
\label{chibetai}
\frac{\partial \Xi ^{i}}{\partial \beta _{i}}=\left( \frac{2}{\beta ^{6}}\left( 2f+3g\right) -\frac{1}{\beta ^{5}}\left( \frac{df}{d\beta }+\frac{dg}{d\beta }\right) \right) B^{2}
\end{equation}

We have to find the functional form of \( f \) and \( g \). This will be achieved
by comparing with the microscopic model to which we will now turn.

\subsection{Computing the interacting potential from the numerical data }

After this rather long interlude, we are now ready to return to the initial
quantities that describe our field at \( t=0. \) Working from the approximate
form of the gap equation, the total energy and the pressure, we will obtain
an appropriate form for the interacting potential, which means that we will
work out the specific form of the functions \( F \) and \( G \) in eq. (\ref{intpot}).
Initially, the total energy density and pressure for the fluid,in the microscopic
model, are given by: 
\begin{equation}
\label{sup57}
T^{00}_{+}=\frac{p_{0}^{2}}{2}+V(\phi _{0})+\rho _{f}+\frac{\partial \Xi ^{0}}{\partial \beta _{0}}
\end{equation}

\begin{equation}
\label{sup58}
T^{ii}_{+}=\frac{p_{0}^{2}}{2}-V(\phi _{0})+p_{f}+\frac{\partial \Xi ^{i}}{\partial \beta _{i}}
\end{equation}

To connect with the microscopic model, we will demand, that 
\begin{eqnarray}
T^{00}_{0} & = & H_{0}\label{sup59} \\
T^{ii}_{0} & = & P_{0}
\end{eqnarray}

with \( H_{0} \) and \( P_{0} \) define in (\ref{H0sim}) and (\ref{P0sim})
respectively. This is a two fluid model. Each fluid possesses a temperature,
or inverse-temperature \( \beta =(k_{B}T)^{-1} \). We thus have here two temperatures
\( T_{c} \) and \( T_{f} \) representing the temperature of each fluid. It
is convenient to work with \( \beta =(\beta _{c}+\beta _{f})/2 \) and \( B=\beta _{c}-\beta _{f} \)
. Defining 
\begin{eqnarray}
D_{H} & \equiv  & H_{0}-\frac{p_{0}^{2}}{2}-V(\phi _{0})-\rho _{f}(0)\label{DH-exact} \\
D_{P} & \equiv  & P_{0}-\frac{p_{0}^{2}}{2}+V(\phi _{0})-p_{f}(0)\label{DP-exact} 
\end{eqnarray}

where \( \rho _{f}(0) \) and \( p_{f}(0) \) are defined in eq. (\ref{rhof})
and (\ref{pf}). As \( \phi _{0}\rightarrow 0, \) numerical simulations indicates
that the \( D \)\'{ }s goes to zero as \( \phi _{0}^{2} \). The functions
\( D_{H} \) and \( D_{P} \) can be approximated with great accuracy as a low
order polynomial (typically of order two or three).

\begin{equation}
\label{DH1}
D_{H}=\sum _{\hat{i}=0}^{imax}c_{i}T_{f}^{i}\phi ^{2}_{0}
\end{equation}

and 
\begin{equation}
\label{DP1}
D_{P}=\sum _{\hat{i}=0}^{imax}d_{i}T_{f}^{i}\phi _{0}^{2}
\end{equation}

\begin{figure}
\par\centering \includegraphics{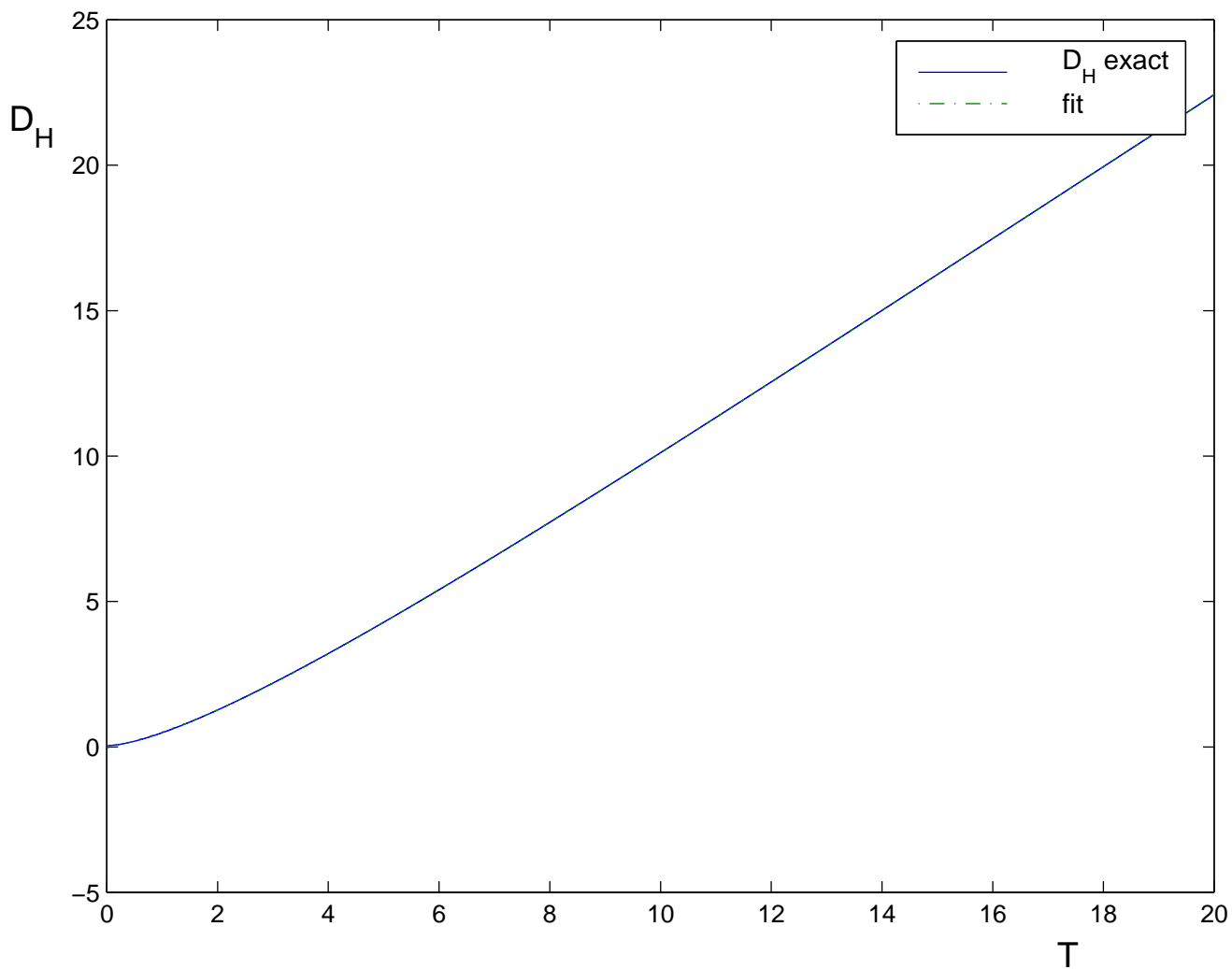}  \par{}

\caption{\label{t00}\protect\protect\( D_{H}\protect \protect \) (eq. (\ref{DH1}))as
a function of \protect\protect\( T\protect \protect \) for fixed \protect\protect\( \phi _{0}=m\protect \protect \)
. The full line is \protect\protect\( D_{H}\protect \protect \) in Eq. (\ref{DH-exact}).
The dotted line represent the least square cubic fit given by Eq. (\ref{DH1})
with \protect\protect\( i_{max}=6.\protect \protect \) }
\end{figure}

\begin{figure}
\par\centering \includegraphics{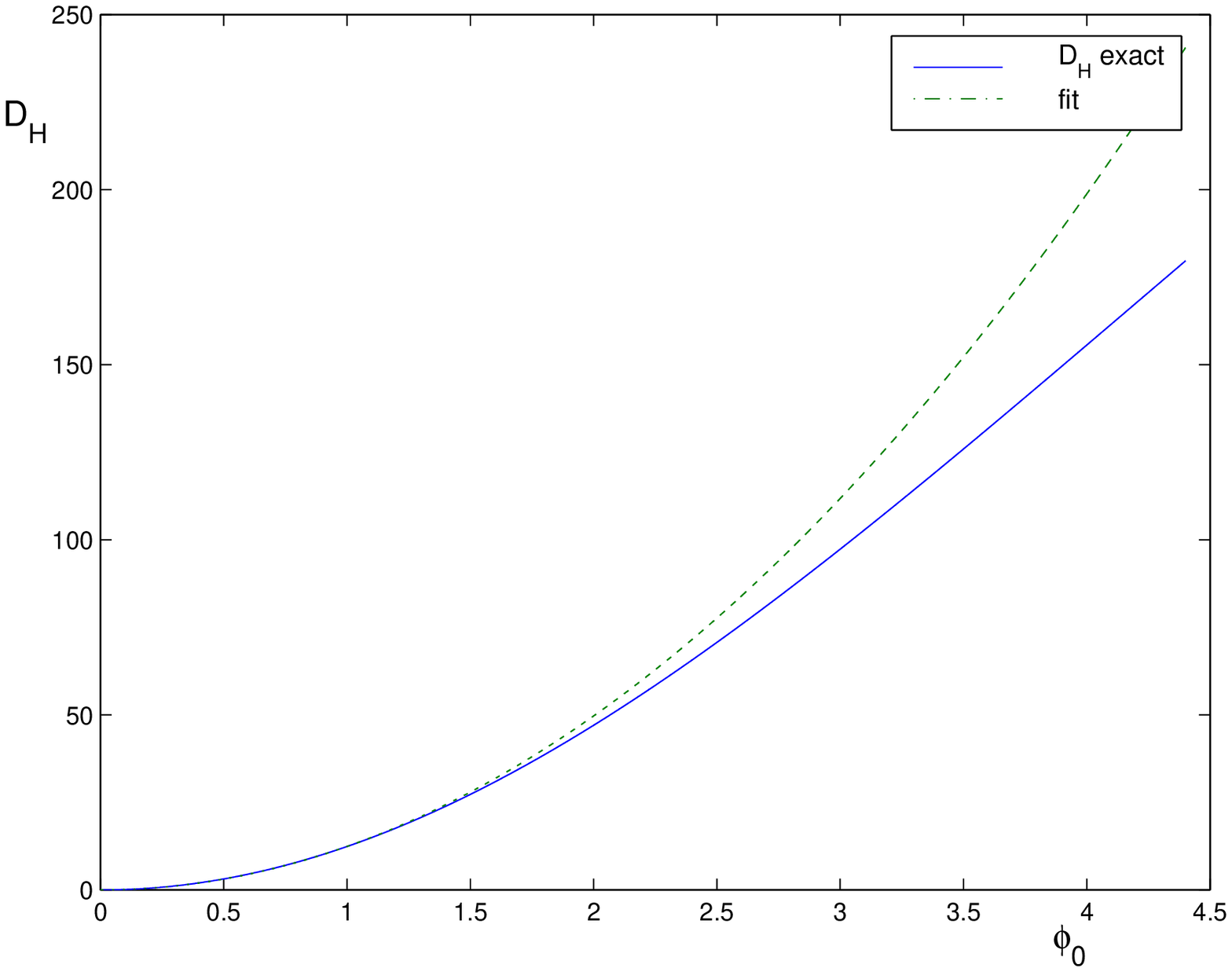}  \par{}

\caption{\label{figure2}Exact \protect\protect\( D_{H}\protect \protect \) ( eq. (\ref{DH1}))
as a function of \protect\protect\( \phi \protect \protect \) at \protect\protect\( T=T_{f}=14\protect \protect \).
The full line is \protect\protect\( D_{H}\protect \protect \) in Eq. (\ref{DH-exact}).
The dotted line represent the quadratic ansatz given by Eq. (\ref{DH1}). }
\end{figure}

\begin{figure}
\par\centering \includegraphics{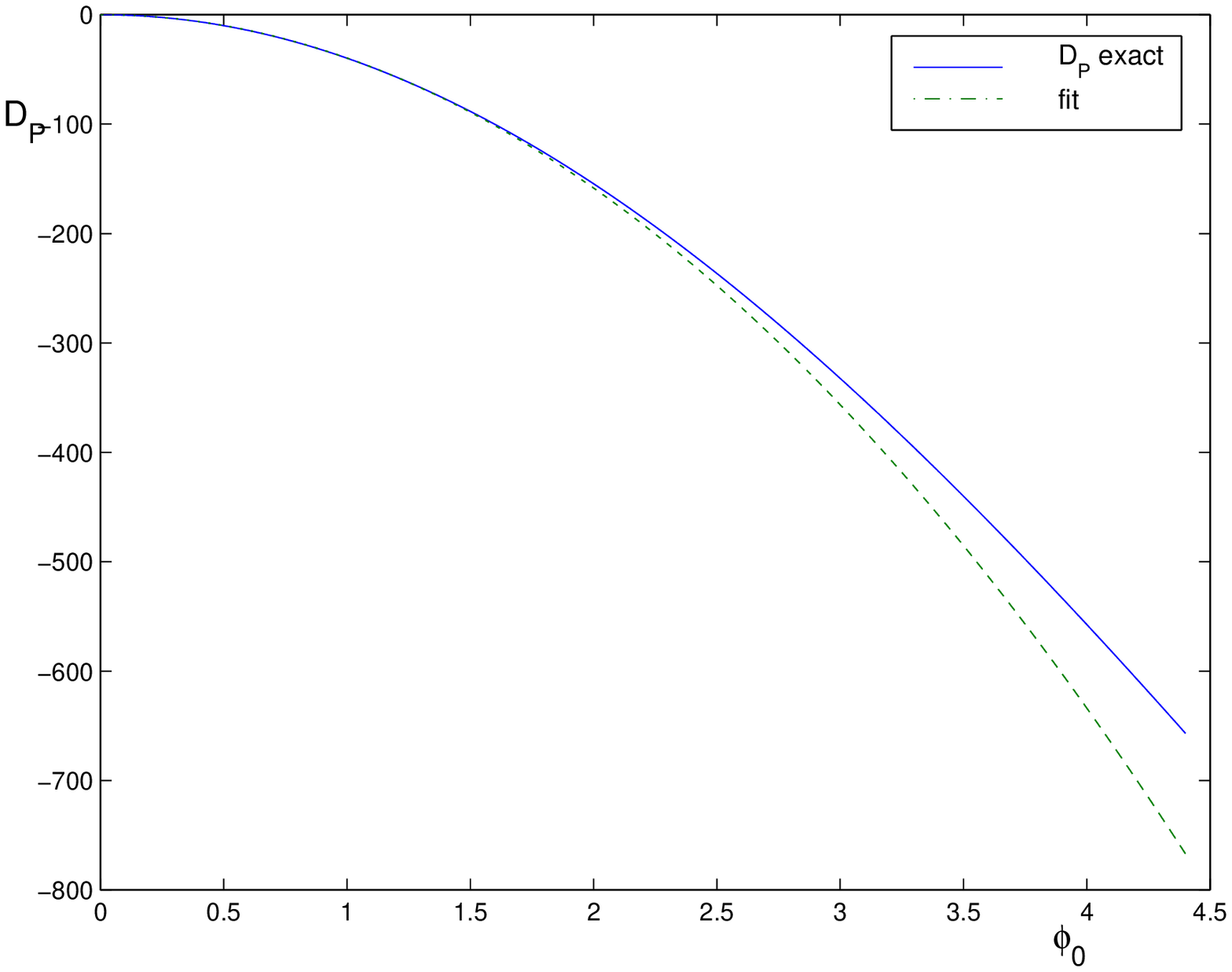}  \par{}

\caption{\label{figure3}Exact \protect\protect\( D_{P}\protect \protect \) (eq. (\ref{DP1}))as
a function of \protect\protect\( \phi \protect \protect \) for \protect\protect\( T=T_{f}=14\protect \protect \).
The full line is \protect\protect\( D_{H}\protect \protect \) in Eq. (\ref{DP-exact}).
The dotted line represent the quadratic ansatz given by Eq. (\ref{DP1}). }
\end{figure}

Both fit were computed to the same order. The coefficients \( c_{i} \) and
\( d_{i} \) can be obtained numerically using a least square fit. Note, for
fixed \( \phi _{0}, \) we have \( t^{00},t^{ii}\rightarrow 0 \) as \( T\rightarrow 0 \).
As an example, we show \( D_{H} \) at fixed \( \phi _{0} \) as a function
of \( T \) (see figure (\ref{t00})). The fit is shown superposed to and is
practically indistinguishable from the exact expression. We also show the exact
\( D_{H} \) and \( D_{P} \) at fixed \( T \) as a function of \( \phi _{0} \)
compared with out quadratic ansatz in figures (\ref{figure2}) and (\ref{figure3}).
The full line shows \( D_{H} \) and \( D_{P} \) defined in Eqs. (\ref{DH-exact})
and (\ref{DP-exact}). The dotted lines are the polynomials approximations (\ref{DH1})
and (\ref{DP1}), respectively.

\vspace{0.3cm}

\subsubsection{Relating \protect\protect\protect\protect\( B_{0}\protect \protect \protect \protect \)
to \protect\protect\protect\protect\( \phi _{0}\protect \protect \protect \protect \)}

In the microscopic model, there is only one temperature \( T_{f} \). However,
the description of the 2-fluids needs two. At equilibrium, we know that there
is only one (uniform) temperature present. That is, at equilibrium, the difference
between the temperature of the fluctuations and of the Klein-Gordon fluid should
be zero. In this limit, the interaction between the two should go (quadratically)
to zero. Turning now to the microscopic formulation, we saw that, as the initial
values of the scalar field goes to zero, the interaction term goes to zero quadratically.
The initial value of the temperature difference is fixed by the condition that
the total energy-momentum tensor of the macroscopic model matches the same quantity
of the microscopic model. As the latter depends on \( \phi _{0} \), this suggests
that \( \phi _{0} \) and the initial temperature difference are linearly related
\begin{equation}
\label{ansatz}
\phi _{0}=\left( \frac{\Lambda }{T_{0}}\right) ^{\alpha /2}k_{B}\left( T_{c0}-T_{f0}\right) 
\end{equation}
 This equations defines the \'{ }field\'{ } temperature \( T_{c0} \) at \( t=0. \)
Following our philosophy of simplicity, we choose to model the proportionality
term via a power law with \( \alpha  \) as parameter. \( \alpha  \) will be
chosen in order to ensure (local) causality. The cutoff, in inverse wavelength
space, has the units of temperature in natural units. Moreover, it makes sense
to have the interaction potential proportional to the cutoff since the fluctuations
are proportional to the cutoff in the microscopic model (recall Eq. (\ref{kphi})
and the paragraph before it). In practice, this means that we will consider
the following relation:

\begin{eqnarray}
\phi _{0}^{2} & = & \left( \frac{\Lambda }{T}\right) ^{\alpha }\left( k_{B}T_{c}-k_{B}T_{f}\right) ^{2}\\
 & = & \left( \frac{\Lambda }{T}\right) ^{\alpha }\frac{B^{2}}{\beta ^{4}}\label{ansatzB} 
\end{eqnarray}

In the last line we evaluated \( \beta _{c},\beta _{f} \) to zero order, that
is, we take \( \beta _{c},\beta _{f}\simeq \beta  \) since we can only seek
a second order term. We thus obtain the following equations: 
\begin{equation}
\label{ansatz1}
D_{H}=\Lambda ^{\alpha }\sum _{\hat{i}=0}^{imax}c_{i}T^{(i+4-\alpha )}B^{2}
\end{equation}
\begin{equation}
\label{ansatz2}
D_{P}=\Lambda ^{\alpha }\sum _{\hat{i}=0}^{imax}d_{i}T^{(i+4-\alpha )}B^{2}
\end{equation}

with the \( c_{i} \) and \( d_{i} \) being constant (but still dependent of
the parameters of the theory, namely \( m \) and \( \lambda  \)). Observe
that by adopting this definition we model well our interacting term in a certain
range of temperature around \( T_{f} \) but we cannot now take the limit \( T_{f}\rightarrow 0 \).
It is convenient to introduce the following definitions 
\begin{equation}
\label{ansatz10}
t^{00}(T)\equiv \Lambda ^{\alpha }\sum _{\hat{i}=0}^{imax}c_{i}T^{(i+4-\alpha )}
\end{equation}

\begin{equation}
\label{ansatz20}
t^{ii}(T)\equiv \Lambda ^{\alpha }\sum _{\hat{i}=0}^{imax}d_{i}T^{(i+4-\alpha )}
\end{equation}

\begin{figure}
\par\centering \includegraphics{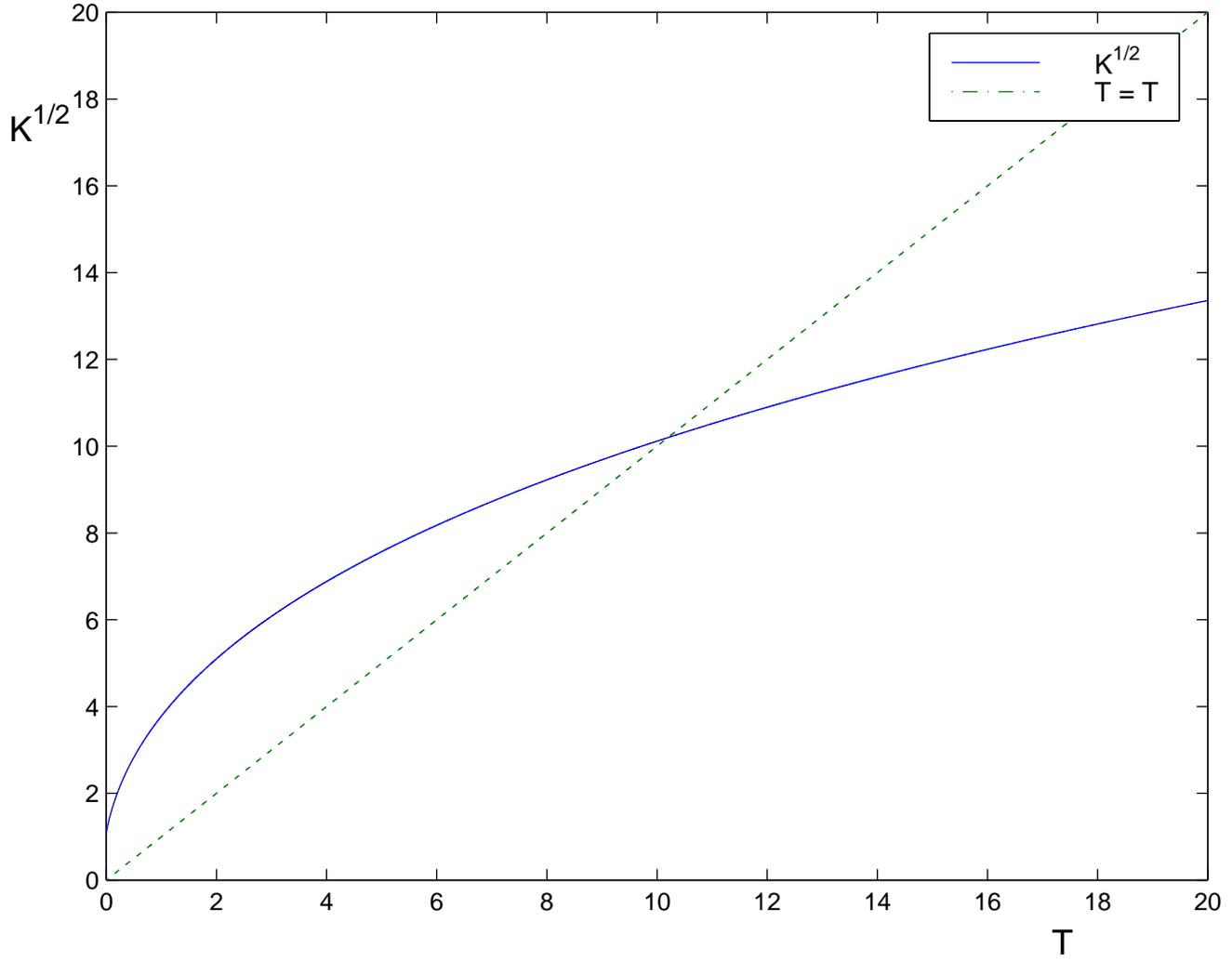}  \par{}

\caption{\label{figure4}\protect\protect\( \sqrt{K(0,0)}\protect \protect \) as a
function of \protect\protect\( T\protect \protect \) . The intersection with
the diagonal dotted line determine \protect\protect\( T_{min}\protect \protect \)
. The value of \protect\protect\( T_{min}\protect \protect \) depends uniquely
on the parameters \protect\protect\( m\protect \protect \) and \protect\protect\( \lambda .\protect \protect \)
With \protect\protect\( m=1.1\protect \protect \) and \protect\protect\( \lambda =15\protect \protect \)
we obtain \protect\protect\( T_{min}\simeq 10.198\protect \protect \)}
\end{figure}

We are now able to compute the interaction potential that we are going to use
in the macroscopic model. However we must pay a price for the crossover from
micro to macro. Indeed, depending on the value of \( \alpha  \), the limit
\( T\rightarrow 0 \) could be ill defined. If we return to the microscopic
model, our limit \( \hbar \rightarrow 0 \) implies that we do not expect our
model to be valid at low temperature. Indeed, we do not expect to obtain valid
result when \( k_{B}T_{f}\ll \omega _{k}(0)=\sqrt{k^{2}+K\left( 0,\phi _{0}\right) } \)
(see Eq. (\ref{nk})). This indicates that we could obtain a minimum temperature
comparing the thermal energy \( k_{B}T_{f} \) to the effective mass at \( \phi _{0}=0. \)
To find \( T_{min} \) we consider \( \sqrt{K\left( 0,0\right) } \) versus
\( T \) (see figure). For value of \( T \) less then \( \sqrt{K(0,0)} \)
then we expect that our ansatz could go awry. We will then define \( T_{min} \)
as the (unique) solution to the equation 
\begin{equation}
\label{Tmin}
\sqrt{K(0,0,T_{min})}-T_{min}=0
\end{equation}

with \( K(0,0) \) the solution of eq.(\ref{kphi}) with \( \phi _{0}=0 \)
(see fig.(\ref{figure4})). With \( m=1.1 \) and \( \lambda =15 \) we obtained
\( T_{min}\simeq 10.198 \)

\vspace{0.3cm}

\subsection{Computation of \protect\protect\protect\protect\( F\protect \protect \protect \protect \)
and \protect\protect\protect\protect\( G\protect \protect \protect \protect \)}

The interaction potential (\ref{intpot}) is defined by two functions of the
mean temperature \( F \) and \( G \), or equivalently the dimensionless \( f \)
and \( g \) defined in (\ref{Ff}) and (\ref{Gg}). Now that we have extracted
what should be the interacting part from the microscopic model at time \( t=0, \)
and made the correspondence with the variables (that is \( T \) and \( B) \)
of the macroscopic model, we are ready to compute explicitly the interaction
potential. Using (\ref{chibeta0}) and (\ref{chibetai}) together with (\ref{ansatz10})
and (\ref{ansatz20}) we obtain: 
\begin{equation}
\label{sup60}
\frac{d^{2}}{d\beta ^{2}}\left( \frac{f+g}{\beta ^{4}}\right) =\tilde{t}^{00}
\end{equation}

\begin{equation}
\label{sup61}
\frac{2}{\beta ^{6}}\left( 2f+3g\right) -\frac{1}{\beta ^{5}}\left( \frac{df}{d\beta }+\frac{dg}{d\beta }\right) =\tilde{t}^{ii}
\end{equation}

with \( \tilde{t}^{00}=t^{00}(\beta ^{-1}) \) and \( \tilde{t}^{ii}=t^{ii}(\beta ^{-1}) \)
as given by (\ref{ansatz10}) and (\ref{ansatz20}). To ease the notation we
will drop the tilde from now on. It is not complicated to turn around these
equations to obtain expressions for \( f \) and \( g \) . As we explained
earlier, we have to set our limit of integration in the range where we believe
our ansatz for \( t^{00} \) and \( t^{ii} \) to be acceptable. Defining \( \beta _{max}=T^{-1}_{min} \)
we obtain: 
\begin{equation}
\label{fmasg}
f+g=\beta ^{4}\int _{\beta }^{\beta _{max}}dx\int _{x}^{\beta _{max}}dy\: t^{00}(y)+a\beta ^{5}+b\beta ^{4}
\end{equation}
 where \( a \) and \( b \) are integration constants. \( b \) in fact will
turn to be irrelevant. We have 
\begin{eqnarray}
t^{ii} & = & \frac{2}{\beta ^{6}}\left( 2f+3g\right) -\frac{1}{\beta ^{5}}\frac{\partial }{\partial \beta }\left\{ \beta ^{4}\int _{\beta }^{\beta _{max}}dx\int _{x}^{\beta _{max}}dy\: t^{00}(y)+a\beta ^{5}+\beta ^{4}b\right\} \\
 & = & \frac{4}{\beta ^{6}}\left\{ \beta ^{4}\int _{\beta }^{\beta _{max}}dx\int _{x}^{\beta _{max}}dy\: t^{00}(y)-g\right\} +\frac{6}{\beta ^{6}}g-\frac{a}{\beta }-\frac{4}{\beta ^{2}}\int _{\beta }^{\beta _{max}}dx\int _{x}^{\beta _{max}}dy\: t^{00}(y)+\frac{1}{\beta }\int _{\beta }^{\beta _{max}}dy\: t^{00}(y)\\
 & = & \frac{2}{\beta ^{6}}g+\frac{1}{\beta }\int ^{\beta _{max}}_{\beta }dy\: t^{00}(y)-\frac{a}{\beta }\label{sup62} 
\end{eqnarray}

leading to 
\begin{equation}
\label{sup63}
g=\frac{1}{2}\beta ^{6}t^{ii}-\frac{1}{2}\beta ^{5}\int _{\beta }^{\beta _{max}}dy\: t^{00}(y)+\frac{1}{2}\beta ^{5}a
\end{equation}

Replacing in the expression for \( f+g \) we immediately find

\begin{equation}
\label{sup64}
f=\beta ^{4}\int _{\beta }^{\beta _{max}}dx\int _{x}^{\beta _{max}}dy\: t^{00}(y)-\frac{1}{2}\beta ^{6}t^{ii}+\frac{1}{2}\beta ^{5}\int _{\beta }^{\beta _{max}}dy\: t^{00}(y)+\frac{1}{2}a\beta ^{5}+b\beta ^{4}
\end{equation}

That is, using (\ref{Ff}) and (\ref{Gg}) 
\begin{equation}
\label{sup65}
F=\int _{\beta }^{\beta _{max}}dx\int _{x}^{\beta _{max}}dy\: t^{00}(y)-\frac{1}{2}\beta ^{2}t^{ii}+\frac{1}{2}\beta \int _{\beta }^{\beta _{max}}dy\: t^{00}(y)+\frac{1}{2}a\beta +b
\end{equation}

and 
\begin{equation}
\label{sup66}
G=\frac{1}{2}t^{ii}-\frac{1}{2\beta }\int _{\beta }^{\beta _{max}}dy\: t^{00}(y)+\frac{a}{2\beta }
\end{equation}

The causality conditions involve derivatives of \( F \) and \( G \) with respect
of their argument \( u=\beta ^{2}=T^{-2}. \) If we denote by a prime a derivative
with respect to \( u \) , we have the following relation: 
\begin{equation}
\label{sup67}
\frac{dF}{du}=-\frac{1}{2}T^{3}\frac{dF}{dT}
\end{equation}

Reverting to our old notation (that is \( t^{00}=t^{00}(T) \) and similarly
with \( t^{ii} \)) we thus obtain 
\begin{equation}
\label{cond1}
F^{\prime }=-\frac{1}{4}T\int _{T_{min}}^{T}dz\frac{1}{z^{2}}t^{00}(z)-\frac{1}{2}t^{ii}(T)-\frac{1}{4}t^{00}(T)+\frac{T}{4}\frac{\partial }{\partial T}t^{ii}(T)+\frac{1}{4}Ta
\end{equation}

\begin{equation}
\label{sup68}
G=\frac{1}{2}t^{ii}(T)-\frac{T}{2}\int _{T_{min}}^{T}dz\frac{1}{z^{2}}t^{00}(z)+\frac{1}{2}Ta
\end{equation}

\begin{equation}
\label{sup69}
G^{\prime }=-\frac{1}{4}T^{3}\frac{\partial }{\partial T}t^{ii}(T)+\frac{1}{4}T^{3}\int _{T_{min}}^{T}\frac{1}{z^{2}}t^{00}(z)\: dy+\frac{1}{4}T^{2}t^{00}(T)-\frac{1}{4}aT^{3}
\end{equation}

with the prime denoting differentiation with respect to \( u=T^{-2} \) and
a lower integration limit \( T_{min}>0 \) was put in to ensure convergence
(this problem is a consequence of the ansatz Eq. (\ref{ansatzB}) as we anticipated
in the previous section).

\subsection{Restrictions imposed by causality and the second law of thermodynamics}

Now we can analyze the restrictions placed on the model by causality considerations.
Let us recall the (local) causality conditions (\cite{ours2}) 
\begin{equation}
\label{condition1}
\frac{dF}{du}>0
\end{equation}

\begin{equation}
\label{condition2}
\frac{dG}{du}+\frac{1}{\beta ^{2}}\left( \frac{dF}{du}+G\right) <0
\end{equation}

\begin{equation}
\label{condition3}
\left( \frac{dG}{du}+\frac{1}{\beta ^{2}}\left( \frac{dF}{du}+G\right) \right) \frac{dF}{du}+\frac{G^{2}}{4\beta ^{2}}<0
\end{equation}

\( \frac{dF}{du} \) was written before (\ref{cond1}). Note that the last two
conditions imply the first. Now 
\begin{equation}
\label{cond2}
\frac{dG}{du}+\frac{1}{\beta ^{2}}\left( \frac{dF}{du}+G\right) =-\frac{T^{3}}{2}\int _{T_{min}}^{T}dz\frac{1}{z^{2}}t^{00}(z)+\frac{1}{2}T^{3}a
\end{equation}

It is simple algebra to write the last condition but the expression is very
messy and not very illuminating. Note that the second condition is verified
whenever \( t^{00}>0 \) which is the case in our microscopic system. Note that
\( b \) does not appear and is therefore irrelevant. The conditions are fulfilled
for a wide range of value of \( \alpha . \) However, we will see later that
the best fit for the model happens with the value of \( \alpha  \) that almost
saturates the conditions, that is that make the conditions almost zero for \( T=T_{f} \)
. The value of \( a \) that allows the best fit with the microscopic theory
was seen to correspond to \( a=0. \)

\begin{figure}
\par\centering \includegraphics{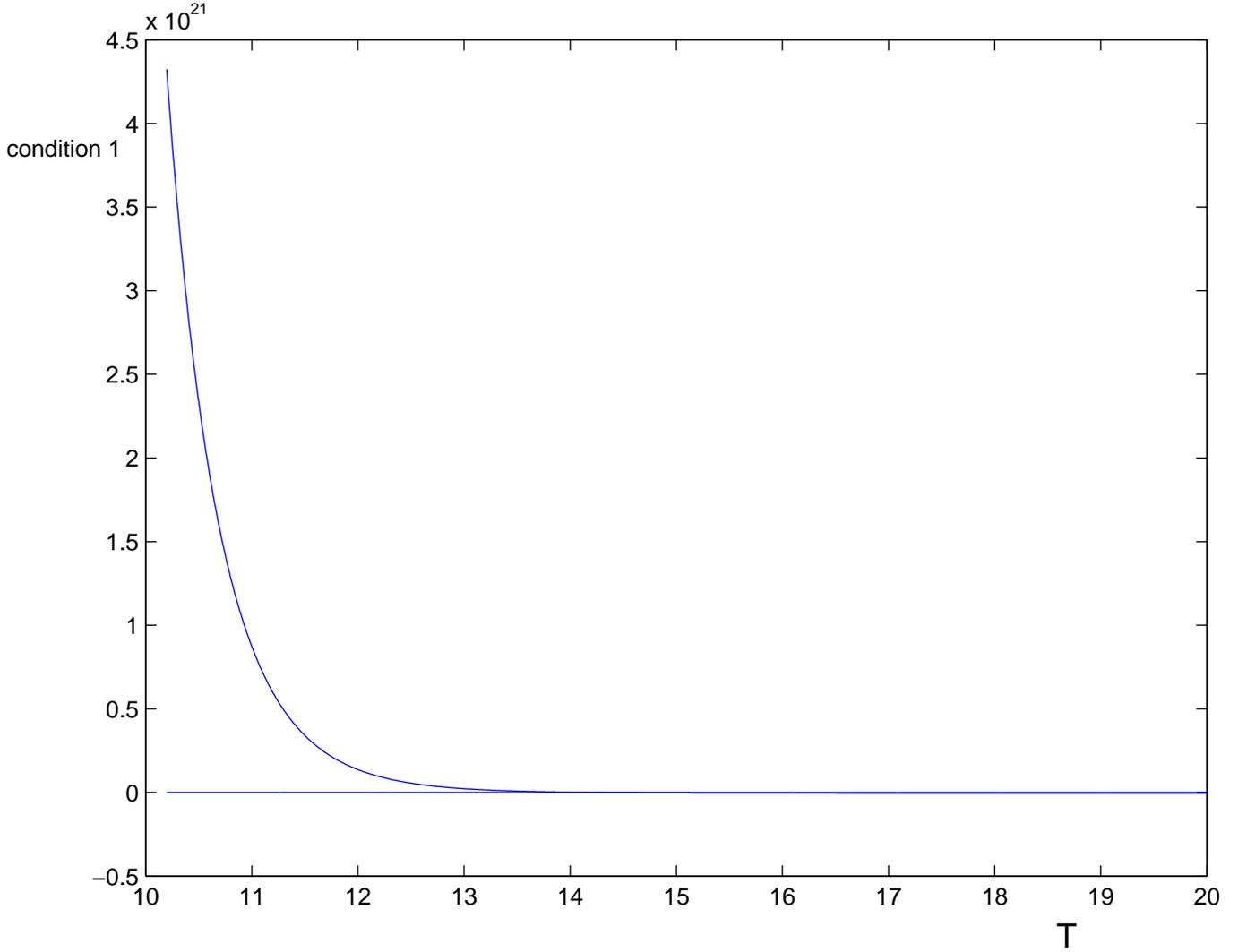}  \par{}

\caption{\label{fig_cond1}The first causality\'{ }s condition, eq. (\ref{condition1}),
as a function of temperature. The allowed range of temperature is given when
the function is positive. The value of \protect\protect\( \alpha =25.9,\protect \protect \)
with \protect\protect\( T_{min}=10.198\protect \protect \) and the integration
constant \protect\protect\( a\protect \protect \) (see eq. (\ref{fmasg}))was
set to zero. Other parameters were \protect\protect\( m=1.1,\protect \protect \) \protect\protect\( \lambda =15\protect \protect \),
the cutoff \protect\protect\( \Lambda =40\protect \protect \) and the initial
conditions \protect\protect\( \phi _{0}=0\protect \protect \) and \protect\protect\( p_{0}=0\protect \protect \)
(see eq. (\ref{ansatz10}) and (\ref{ansatz20}) ). }
\end{figure}

\begin{figure}
\par\centering \includegraphics{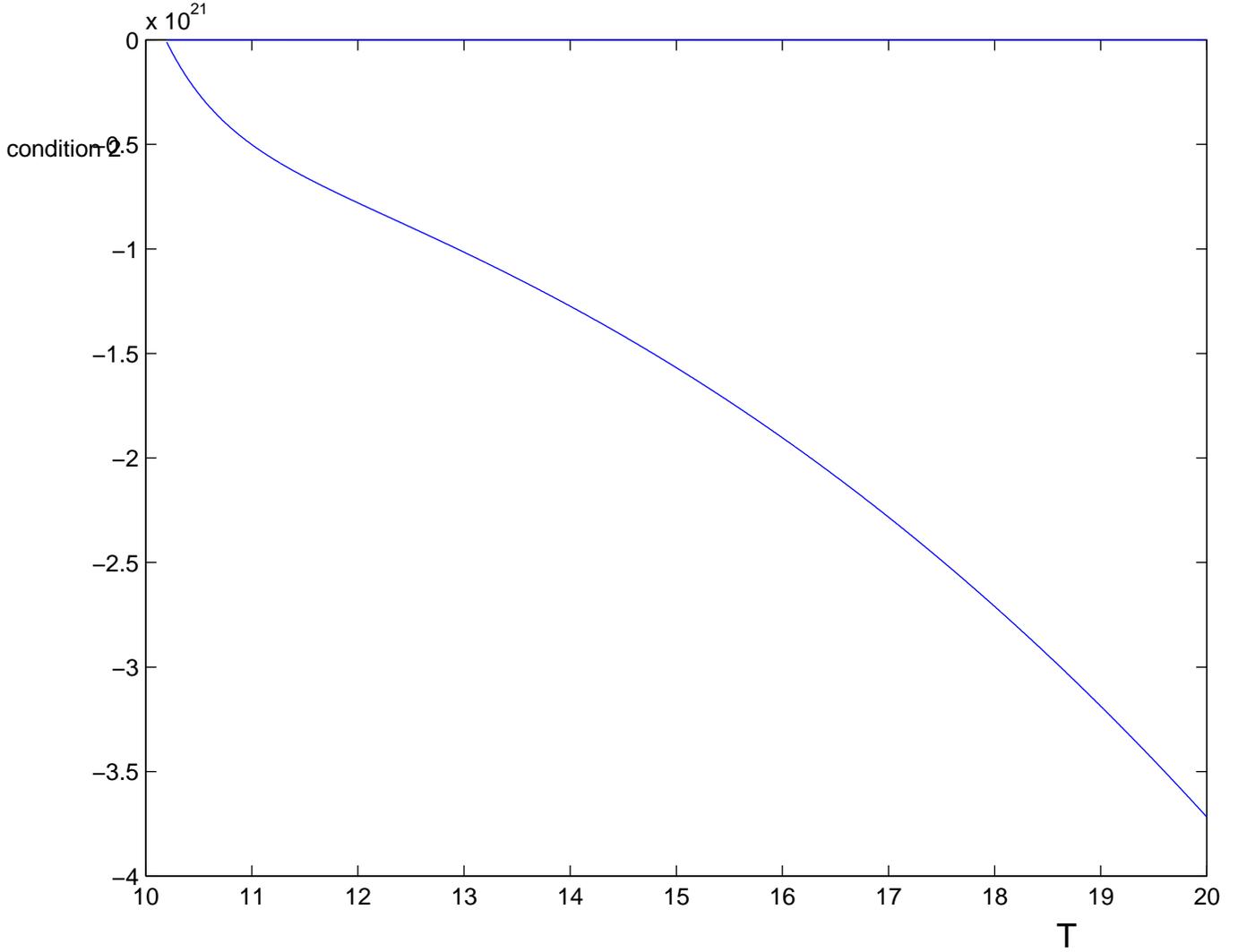}  \par{}

\caption{\label{fig_cond2}The second causality\'{ }s condition, eq. (\ref{condition2}),
as a function of temperature. See the caption of fig. (\ref{fig_cond1}) for
details. }
\end{figure}

\begin{figure}
\par\centering \includegraphics{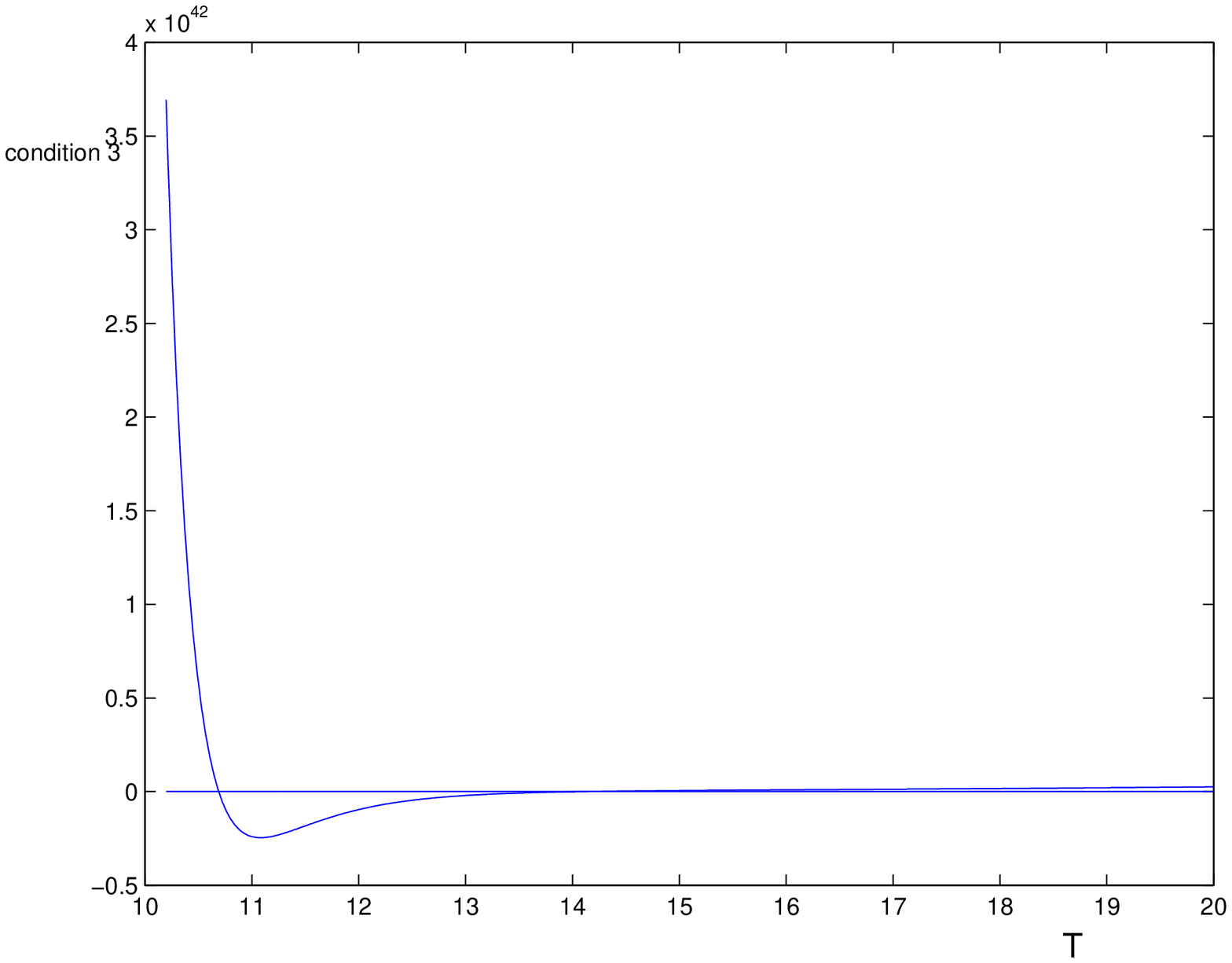}  \par{}

\caption{\label{fig_cond3}The third causality\'{ }s condition, eq. (\ref{condition3}).
See the caption of fig. (\ref{fig_cond1}) for details. }
\end{figure}

The first conditions (\ref{condition1}) is shown in figure (\ref{fig_cond1}),
the second condition (\ref{condition2}) in figure (\ref{fig_cond2}) and the
third (\ref{condition3}) in figure (\ref{fig_cond3}). All are plotted as a
function of \( T \), with the choice \( \alpha =25.9 \). (with \( m=1.1, \)
\( \phi _{0}=2*m, \) \( p_{0}=0, \) \( \Lambda =40 \), \( \lambda =15 \)
and \( T_{min}=\sim 10.198. \) as given by figure (\ref{figure4})).

Usually the conditions are fulfilled within a range of value for \( \alpha  \)
that depends on the value of \( T_{min.} \). \( \alpha  \)is greater then
\( \sim 6 \) with \( T_{min}\simeq 0 \) but the minimum value for \( \alpha  \)
grows with increasing \( T_{min.} \) Therefore, within a finite range of temperature
( from approximately \( T=10.6 \) to \( T=T_{f}=14 \) in the case of figures
(\ref{fig_cond1}) to (\ref{fig_cond3})), the three conditions are met thus
ensuring causality and stability.

\section{Comparison between the microscopic and the macroscopic model}

\subsection{Matching the dynamics}

To compare the two models, we have to choose parameters in a consistent way
to ensure that we are comparing effectively the same physical system. Some parameters
are trivially chosen since we assume that the mass of the particle \( m \)
and the coupling constant \( \lambda  \) are exactly the same in both models.
We also have already taken for granted that the temperature \( k_{B}T_{f} \)
in the microscopic model is equal to \( \beta ^{-1}_{f0}=\left( \beta _{0}-B_{0}/2\right) ^{-1}. \)
The quantities that should be equal, at least initially are the total energy
density, the total pressure and \( K(0,\phi _{0}). \) Note that the initial
condition for the field and its conjugate momentum should also be the same in
both theories. Our microscopic system conserves both the total energy density
and the entropy. However, since the macroscopic system is a coarse grained version
of the full system, one could expect some generation of entropy in the macroscopic
model. To describe this, we will allow a nonzero entropy production. The entropy
current is defined as follows

\begin{equation}
\label{sup70}
S^{a}=\chi ^{a}-\beta _{(c)b}T^{ab}-\beta _{(f)b}T^{ab}-\xi j^{a}
\end{equation}

As we said earlier, the entropy production is computed to be 
\begin{equation}
\label{nablaS}
\nabla _{a}S^{a}=-B_{a}I^{a}-\xi \Delta 
\end{equation}

where the equations of motion (\ref{nueva}), (\ref{nueva2}) and (\ref{nueva3})
have been used. Defining 
\begin{eqnarray}
I^{a} & = & D\xi \beta ^{a}+EB^{a}+F\beta _{b}B^{b}\beta ^{a}\label{sup71} \\
\Delta  & = & A\xi +C\beta _{a}B^{a}
\end{eqnarray}

We can rewrite (\ref{nablaS}) in the following manner 
\begin{equation}
\label{sup72}
\nabla _{a}S^{a}=-(A\xi +C\beta _{a}B^{a})\xi -\left( D\xi \beta ^{a}+EB^{a}+F\beta _{b}B^{b}\beta ^{a}\right) B_{a}
\end{equation}
 If one demands \( \nabla _{a}S^{a}=0 \) then \( A=E=F=\left( C+D\right) =0 \).
Relaxing this restrictions to allow some entropy production will introduce more
terms in the equations of motion. More specifically, we use the following ansatz
for the \( \Delta  \) and \( I^{a} \) terms: 
\begin{eqnarray}
\Delta  & = & -A\xi -C\beta B\label{sup73} \\
I^{0} & = & -C\xi \beta 
\end{eqnarray}

That is, 
\begin{equation}
\label{deltaMay}
\Delta =-C\beta B-Ap\left( \frac{1}{T}+\frac{1}{2}B\right) 
\end{equation}
\begin{equation}
\label{Izero}
I^{0}=C\left( \frac{1}{T}+\frac{1}{2}B\right) \frac{1}{T}p
\end{equation}

where \( p \) is the momentum conjugate to \( \phi  \) and \( C \) has units
of \( T^{5} \) and \( A \) of \( T^{2}. \) We choose to model this factor
as follows

\begin{eqnarray}
C & = & C_{0}T^{4}\phi \\
A & = & A_{0}\phi ^{2}\label{A0} 
\end{eqnarray}

Recall that \( \phi  \) has units of temperature in natural units, that we
adopt in this work. The \( \phi  \) dependence is to enforce the fixed point
\( \phi _{c}=0 \). \( A_{0} \) is the entropy production term. Entropy production
will be linked with the decay of the scalar field since, as a consequence of
the conservation of energy, this decay will imply a growth of the energy density
of the fluctuations. Without it, the amplitude of the field stays constant in
time. The value of \( C_{0} \) is chosen by demanding that the initial value
of the left-hand side of the equation for \( \dot{p} \) is the same for both
theories. We thus obtain: 
\begin{equation}
\label{sup74}
C_{0}=\left[ K(0,\phi _{0})-m^{2}-\frac{\lambda }{2}\phi ^{2}_{0}+A_{0}\phi _{0}\left( \frac{1}{T_{0}}+\frac{1}{2}B_{0}\right) p_{0}\right] \frac{1}{T_{0}^{3}B_{0}}
\end{equation}
 We can now return to the equations (\ref{nueva}) to (\ref{old17}) and write
them as a function of our variables. In order to integrate them numerically,
it is convenient to rewrite them in the following form:

\begin{eqnarray}
\dot{\phi } & = & p\\
\dot{p} & = & -V^{\prime }(\phi )-\Delta \\
\dot{T} & = & -\Delta \, p\, f_{1}(T,B)-2I^{0}f_{2}(T,B)\label{sup75} \\
\dot{B} & = & \Delta \, p\, f_{3}(T,B)+2I^{0}f_{4}(T)
\end{eqnarray}
 where \( \Delta  \) and \( I^{0} \) are defined en eq. (\ref{deltaMay})
and (\ref{Izero}). The functions \( f_{i}(\beta ) \) are quite complicated
and are obtained upon substitution of the interacting potential in the equations
of motions. Reordering to display the equations in a form suitable for numerical
integrations, 
\begin{equation}
\label{sup76}
f_{1}=\frac{g_{4}-g_{2}}{g_{1}g_{4}-g_{2}g_{3}}
\end{equation}

\begin{equation}
\label{sup77}
f_{2}=\frac{g_{2}}{g_{1}g_{4}-g_{2}g_{3}}
\end{equation}

\begin{equation}
\label{sup78}
f_{3}=\frac{g_{1}-g_{3}}{g_{1}g_{4}-g_{2}g_{3}}
\end{equation}

\begin{equation}
\label{sup79}
f_{4}=\frac{g_{1}}{g_{1}g_{4}-g_{2}g_{3}}
\end{equation}

where 
\begin{equation}
\label{sup80}
g_{1}=\frac{1}{T^{2}}T_{f}^{2}\rho _{q}^{\prime }+\tau _{1}^{\prime }(T)B^{2}
\end{equation}

\begin{equation}
\label{sup81}
g_{2}=\frac{1}{2}T_{f}^{2}\rho _{q}^{\prime }+2\tau _{1}(T)B
\end{equation}

\begin{equation}
\label{sup82}
g_{3}=-\frac{1}{T^{2}}T_{f}^{2}\rho _{q}^{\prime }+\sigma _{1}^{\prime }(T)B
\end{equation}

\begin{equation}
\label{sup83}
g_{4}=-\frac{1}{2}T_{f}^{2}\rho _{q}^{\prime }+\sigma _{1}(T)
\end{equation}

with \( \rho ^{\prime }_{q} \) means that the derivative is taken with respect
to \( T_{f} \)

\begin{equation}
\label{sup84}
\tau _{1}(T)=\Lambda ^{\alpha }\sum _{i=0}^{imax}c_{i}T^{i+4-\alpha }
\end{equation}

\begin{equation}
\label{sup85}
\sigma _{1}(T)=4\int _{T_{min}}^{T}\frac{1}{z^{2}}\tau _{1}(z)dz
\end{equation}

with the \( c_{i} \) obtained by least square-fitting as explained in section
2.5.1. The integration constant \( a \) that appears in eq.(\ref{fmasg}) has
been set to zero, as discussed above.

\subsection{Results}

We are interested in describing the scalar field and also to be able to predict
the temperature of the fluctuations. We do not expect our coarse-grained model
to match exactly the microscopic one but at least we should be able to follow
the decay of the inflaton and the temperature (which means that the density
of the fluctuations should be accurately predicted). Of the various parameters
that appears in our theory, almost all are fixed by the initials conditions
or the parameters of the theory. \( T_{min} \) is fixed by Eq. (\ref{Tmin})
and depends only on \( m \) and \( \lambda  \). \( C_{0} \) by Eq. (\ref{sup74}).
Finally the constant of integration \( a \) was set to zero. There are two
parameters that can be adjusted: \( A_{0} \) (see eq.(\ref{A0})) and \( \alpha  \).
In both cases, these parameters are not fixed by the initial conditions but
are chosen to match some behavior of the scalar field, namely, to emulate the
decay of the inflaton field in the case of \( A_{0} \) and to match the phase
in the case of \( \alpha  \). However, once fixed, they cannot be changed when
another set of initial values is chosen. In the examples shown below, \( A_{0}=0.9. \)
and \( \alpha =25.9 \). Moreover, as we have seen before, the value of \( \alpha  \)
is the one that almost saturates the causality conditions.

We choose the following set of parameters for the numerical simulations that
we present in the article: \( m=1.1 \), \( \lambda =15 \) (see eq. (\ref{pot})),
\( T_{f}=14 \) (see eq(\ref{Tf})), The high value of \( \lambda  \) is to
ensure that the nonlinear effects are important. The cutoff \( \Lambda =40 \)
(see (\ref{cutoff})) with a spacing \( \Delta _{k}=.05 \) (see (\ref{old12}))
which gives a very good approximation of the integrals by Riemannian sums. This
implies that we work with \( 400 \) complex modes in the microscopical model.
This was integrated in Fortran using the Bulirsch-Stoer method as implemented
in the Numerical Recipes \cite{numerical}. A simple Runge-Kutta adaptive step
algorithm was used to compute the equations system for the macroscopic model.
To compare we show the homogeneous mode of the scalar field of the micro version
against the scalar field in the fluid model.

\begin{figure}
{\par\centering \includegraphics{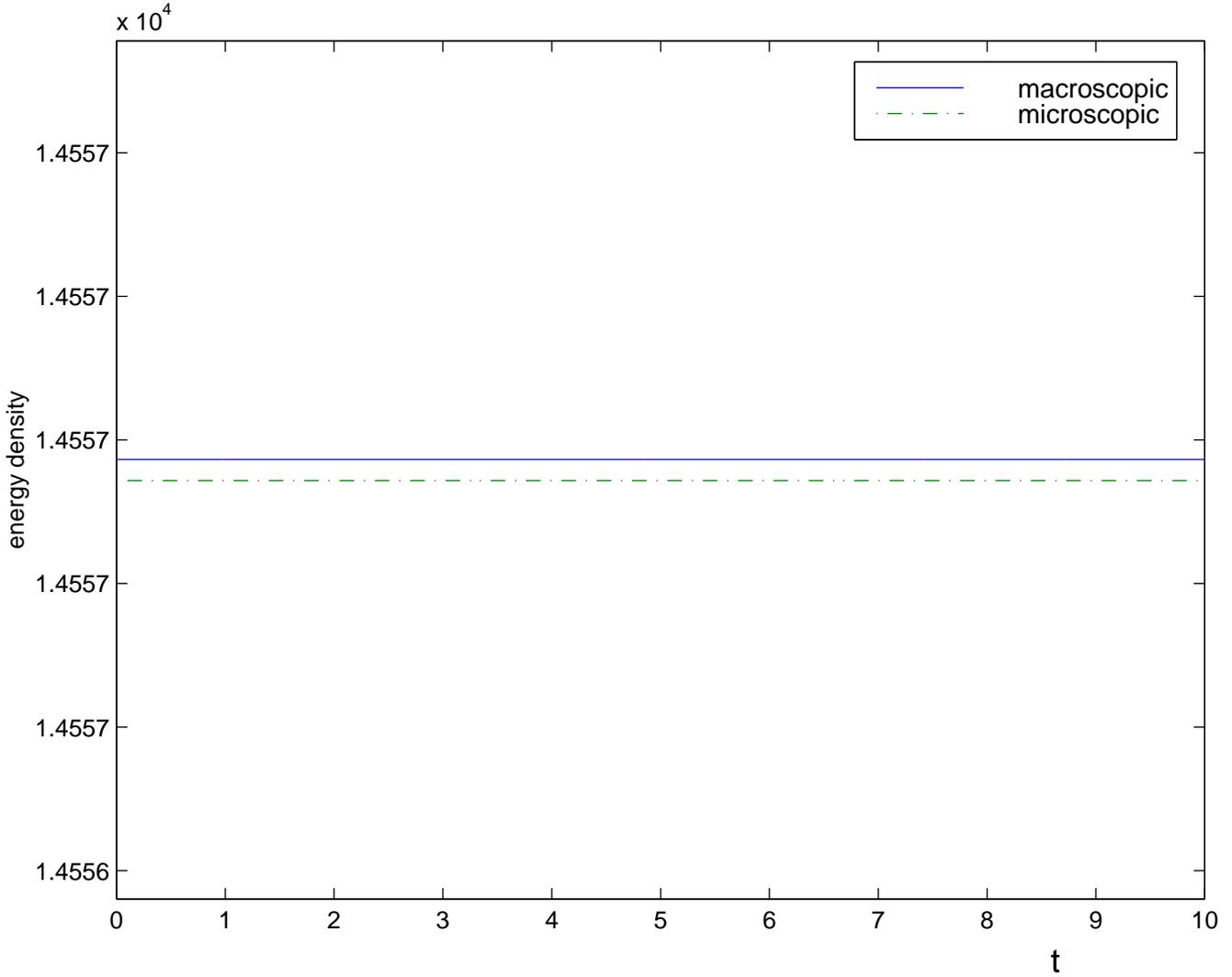} \par}

\caption{\label{energy}Energy density as a function of time for the case \protect\protect\( \phi _{0}=m.\protect \protect \)
The solid line is the macroscopic model and the dash line correspond to the
microscopic model. The slight difference came from the error in the least-square
when we compute the energy density of the fluctuations,\protect\protect\( \rho _{f}\protect \protect \),
and the i nteraction energy, \protect\protect\( t^{00}B^{2}\protect \protect \)}
\end{figure}

Our objective is to demonstrate how the macroscopic model captures the essential
features of the microscopic one. In order to do so we display a number of graphics
that compare the scalar field and its time derivative in the two models. The
initial values are \( \phi _{0}=f*m \) and \( p_{0}=0 \). For the first run,
we set \( f=1 \) and the scalar field is given as a function of time. In this
case, the scalar field by itself is not very different from the one given by
the Klein-Gordon equation with \( m_{eff}^{2}=K(0,0). \) The parameter \( \alpha =25.9 \)
( see (\ref{ansatz})), the value that almost saturated the the causality conditions
(see (\ref{condition1}),(\ref{condition2}) and (\ref{condition3})) , with
\( T_{min}=10.198 \), eq (\ref{Tmin}) . In figure (\ref{energy}), we show
the energy density. Note that in both cases, this is conserved up to numerical
accuracy, as it should be. The slight difference between the total energy density
in the two models is generated by the error inherent to the least square fit
approximation for \( \rho _{f} \) (almost all the energy density came from
the fluctuations).

\begin{figure}
\par\centering \includegraphics{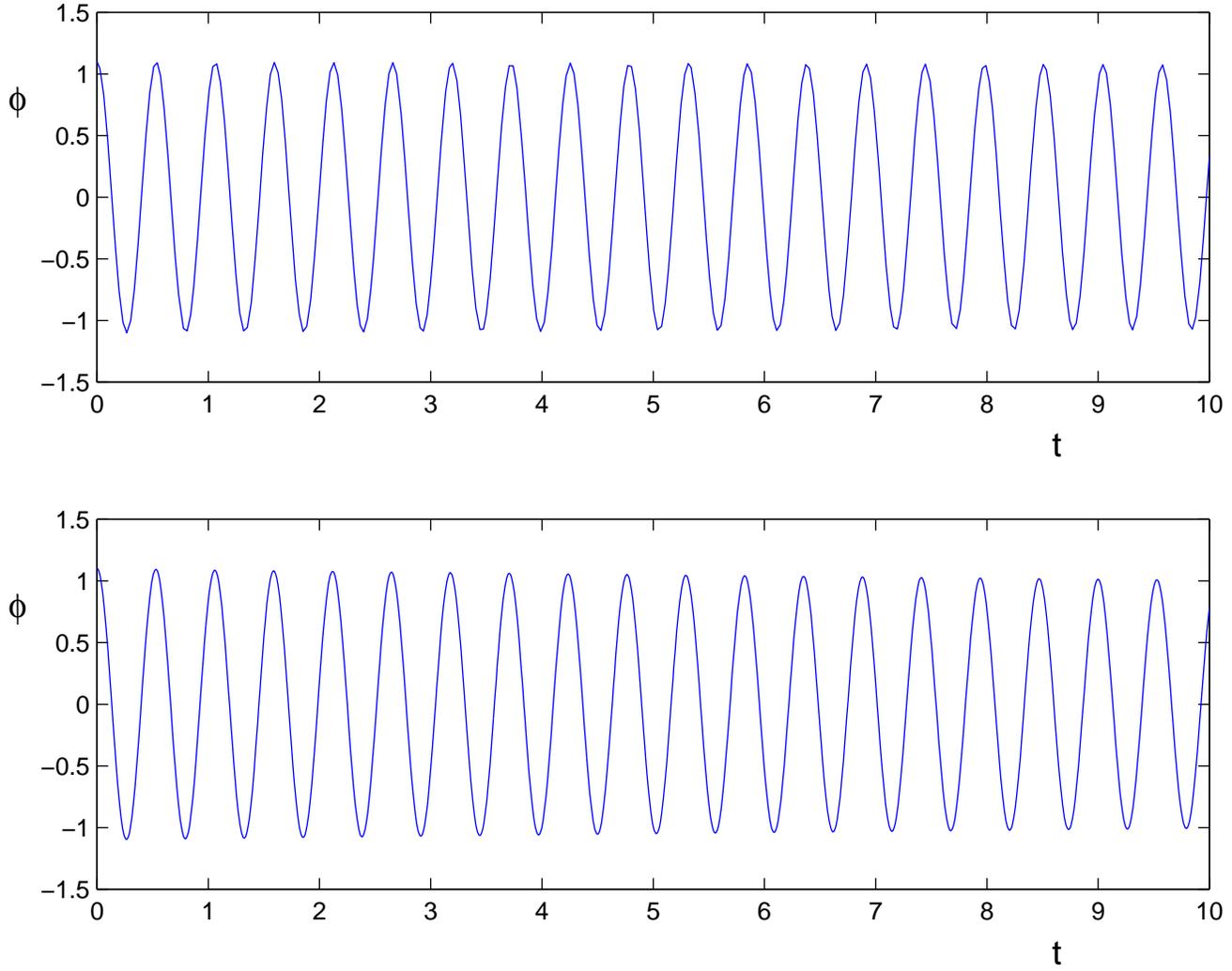}  \par{}

\caption{\label{fig2}The scalar field shown as a function of time. The upper plot correspond
to the microscopic theory and the lower plot to the macroscopic theory. The
initial value for \protect\protect\( \phi _{0}=m\protect \protect \), the parameter
\protect\protect\( A_{0}=0.9,\protect \protect \) the parameter \protect\protect\( a=0\protect \protect \)
and \protect\protect\( \alpha =25.9.\protect \protect \) }
\end{figure}

\begin{figure}
\par\centering \includegraphics{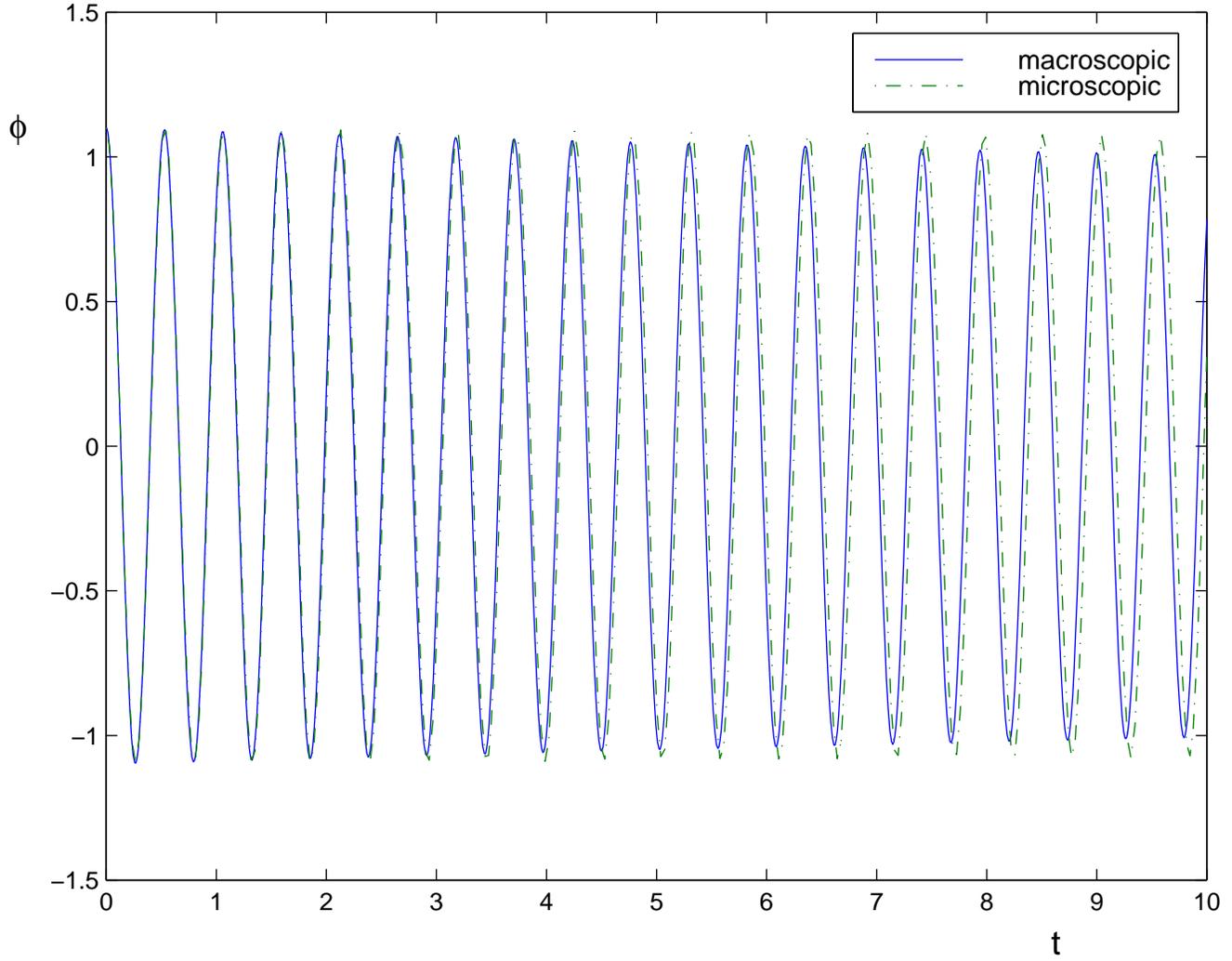}  \par{}

\caption{\label{fig2b}The scalar field of fig.(\ref{fig2}) superimposed. The solid
line represents the macroscopic case and the dotted line the microscopic theory. }
\end{figure}

\begin{figure}
\par\centering \includegraphics{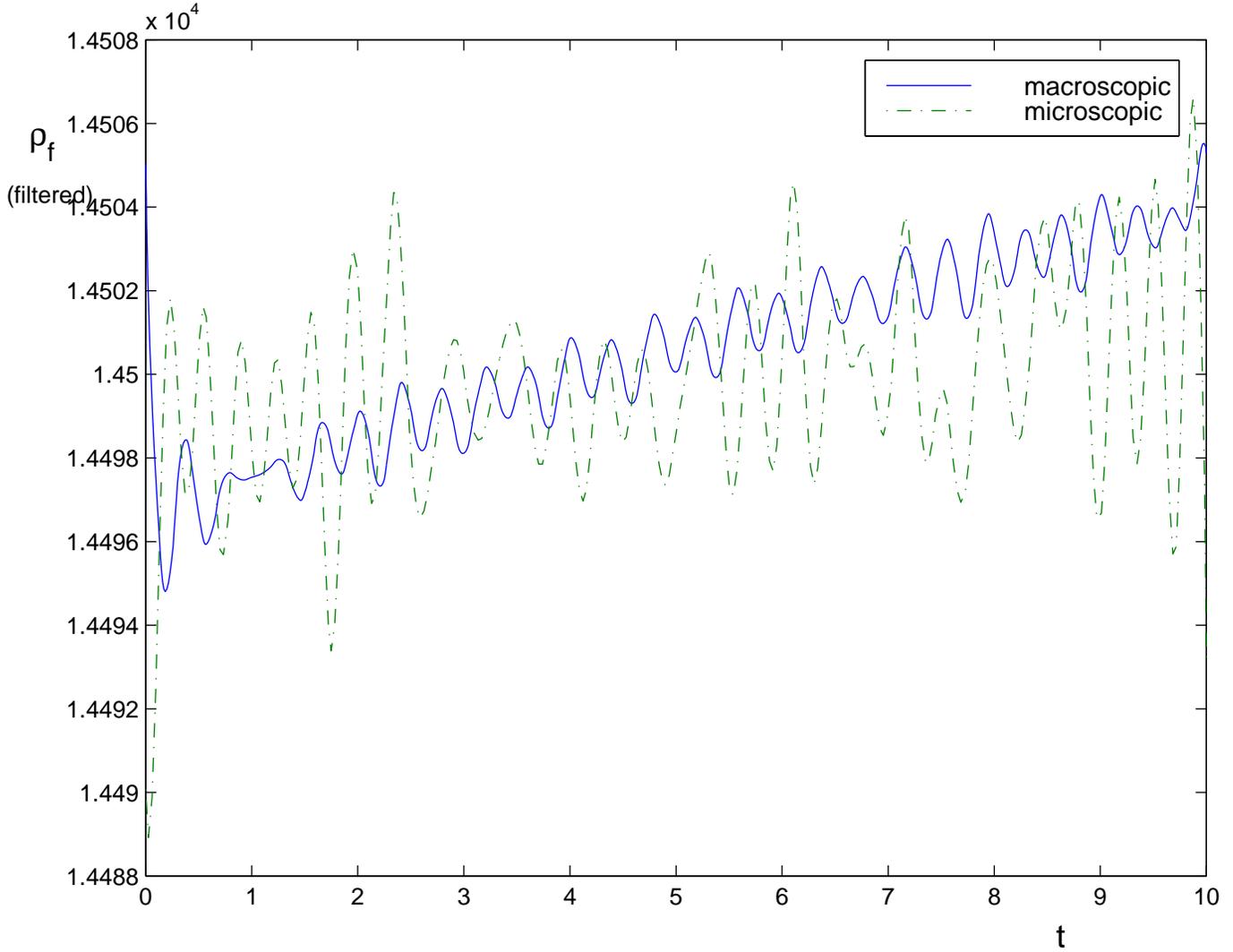}  \par{}

\caption{\label{energyF1}Mean value of the energy density of the fluctuations. The
solid line represent the macroscopic theory given by eq. (\ref{rhomacro})and
the dotted line the microscopic theory given by eq. (\ref{rhomicro}). }
\end{figure}

In figure (\ref{fig2}), the scalar field is depicted as a function of time
for both models. For these values, they are almost equal up to numerical accuracy.
To ease the comparison, the two graphs are shown superimposed in figure (\ref{fig2b}).
In figure (\ref{energyF1}), the energy density of the fluctuations for both
models is also shown. We recall that 
\begin{equation}
\label{rhomacro}
\rho ^{macro}_{f}(t)=\sum _{i=0}^{i_{max}}e_{i}T_{f}^{i}(t)
\end{equation}

where the \( e_{i} \) are obtained by least square fit of eq. (\ref{rhof})
and 
\begin{equation}
\label{sup86}
T_{f}=\left( \frac{1}{T}-\frac{1}{2}B\right) ^{-1}
\end{equation}

and 
\begin{equation}
\label{rhomicro}
\rho _{f}^{micro}(t)=T^{00}_{+}-\frac{1}{2}p^{2}(t)-V(\phi (t))
\end{equation}
 At \( t=0 \), the two should be equal since the interaction energy is null
initially. Note that \( \rho _{f} \) is a function of \( T_{f} \). To make
the reading easier, the output in both cases was processed with a high frequency
filter to get rid of the high frequency and to show the general trend. In the
microscopic model, we define the energy density subtracting from the total energy
density the energy density of the scalar field.

\begin{figure}
\par\centering \includegraphics{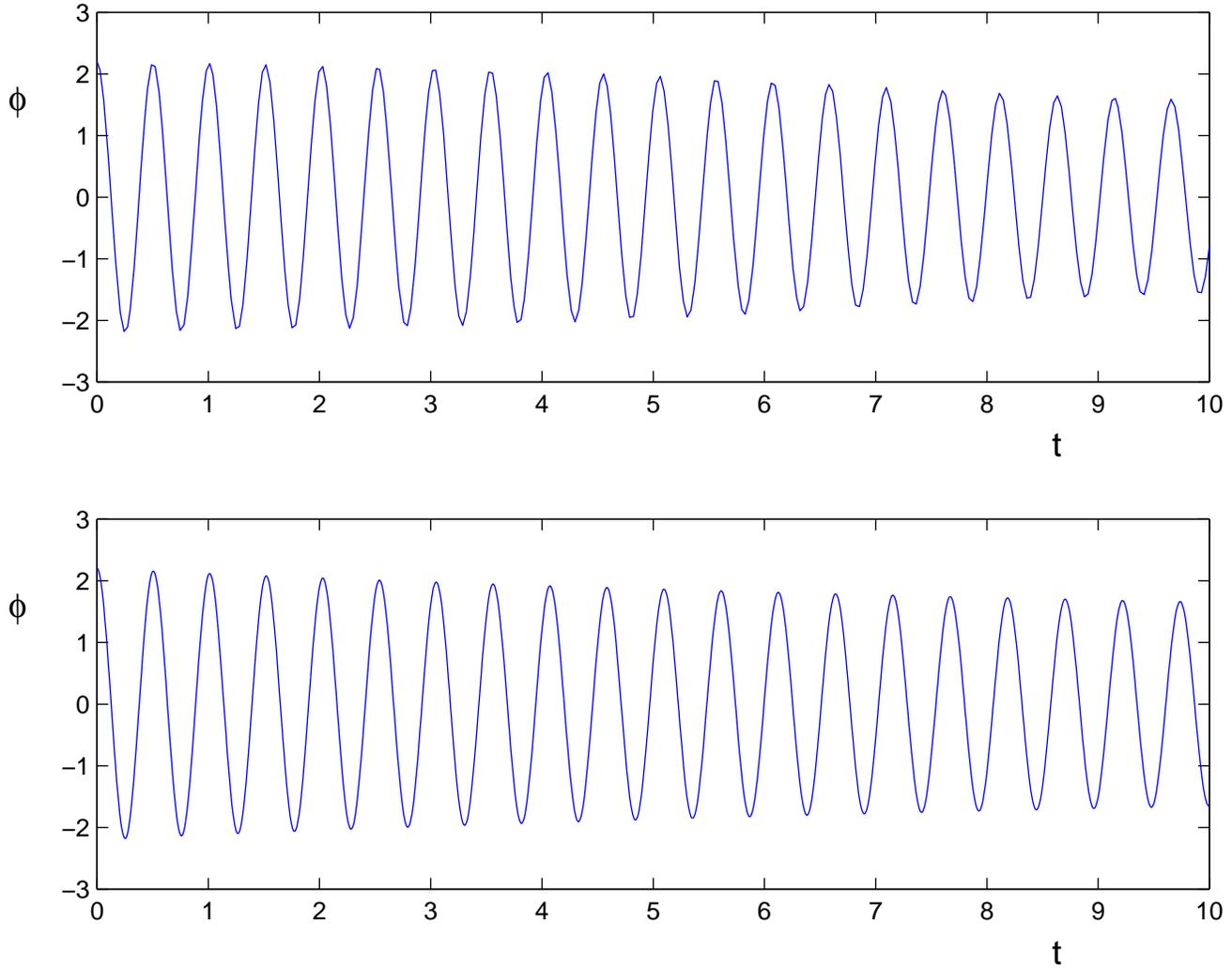}  \par{}

\caption{\label{fig3} The scalar field depicted as a function of time in the case \protect\protect\( \phi _{0}=2m.\protect \protect \)
The upper graphics represents the macroscopic case and the lower one the microscopic
case. All parameters where kept to the same value as in fig. (\ref{fig2}) namely
\protect\protect\( A_{0}=0.9\protect \protect \), \protect\protect\( a=0\protect \protect \)
and \protect\protect\( \alpha =25.9.\protect \protect \) }
\end{figure}

\begin{figure}
\par\centering \includegraphics{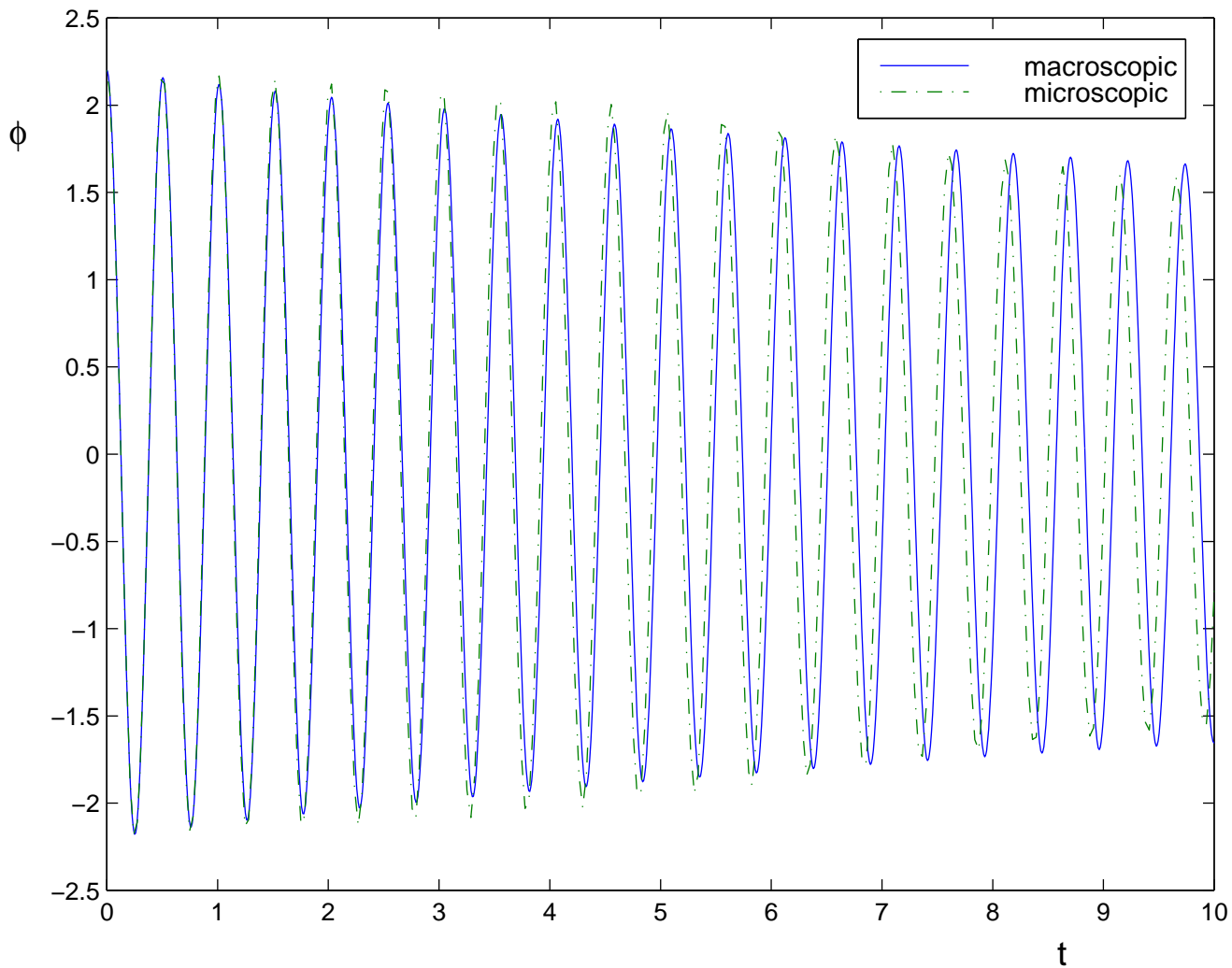}  \par{}

\caption{\label{fig3b}Same as fig.(\ref{fig3}) but shown together. The solid line
is the macroscopic theory and the dotted line is the microscopic theory. }
\end{figure}

\begin{figure}
\par\centering \includegraphics{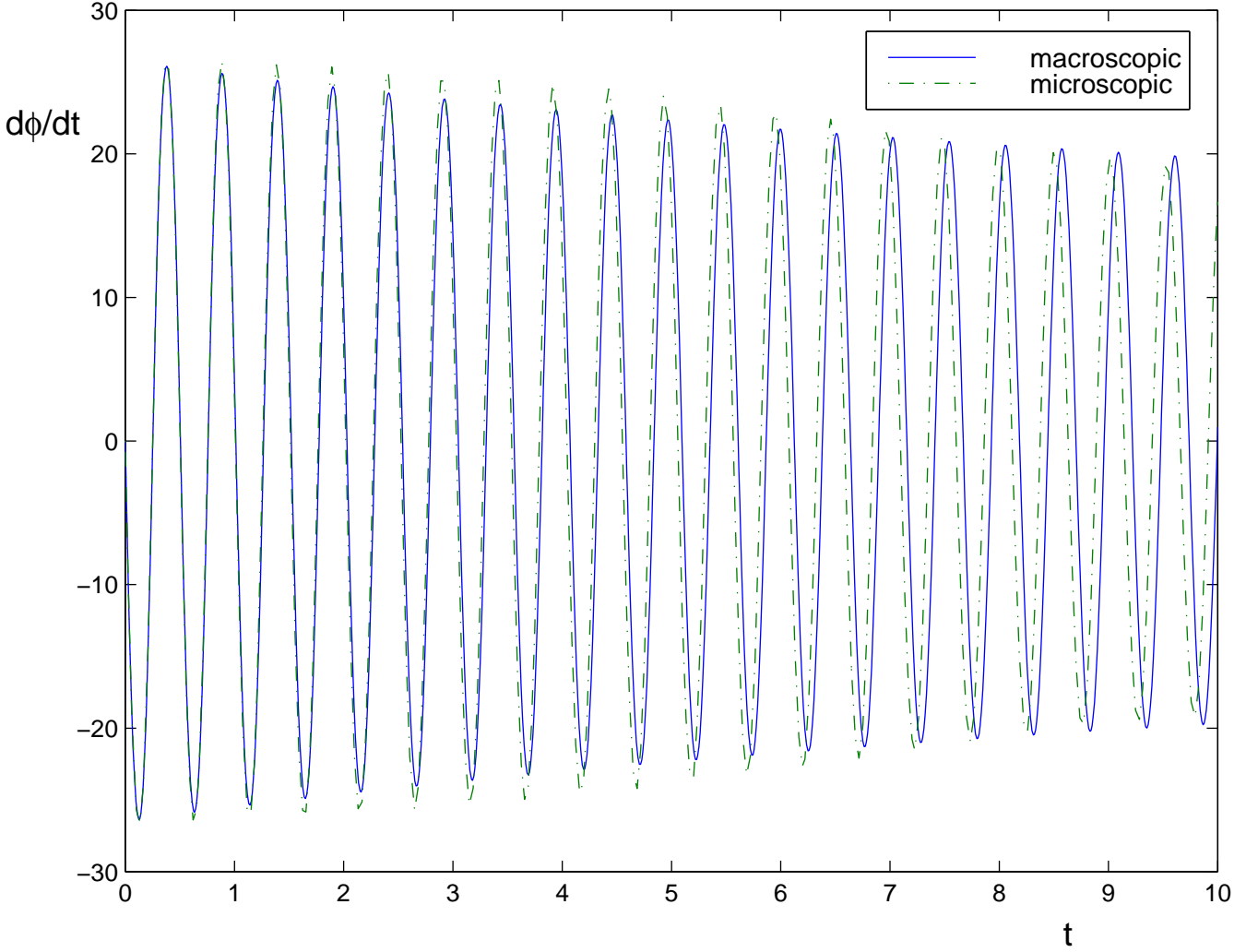}  \par{}

\caption{\label{phidot}The same initials conditions as in fig.(\ref{fig3}) but the
time derivative of the scalar field is shown as a function of time. Solid line
is the macroscopic theory and the dotted line is the microscopic theory. }
\end{figure}

\begin{figure}
\par\centering \includegraphics{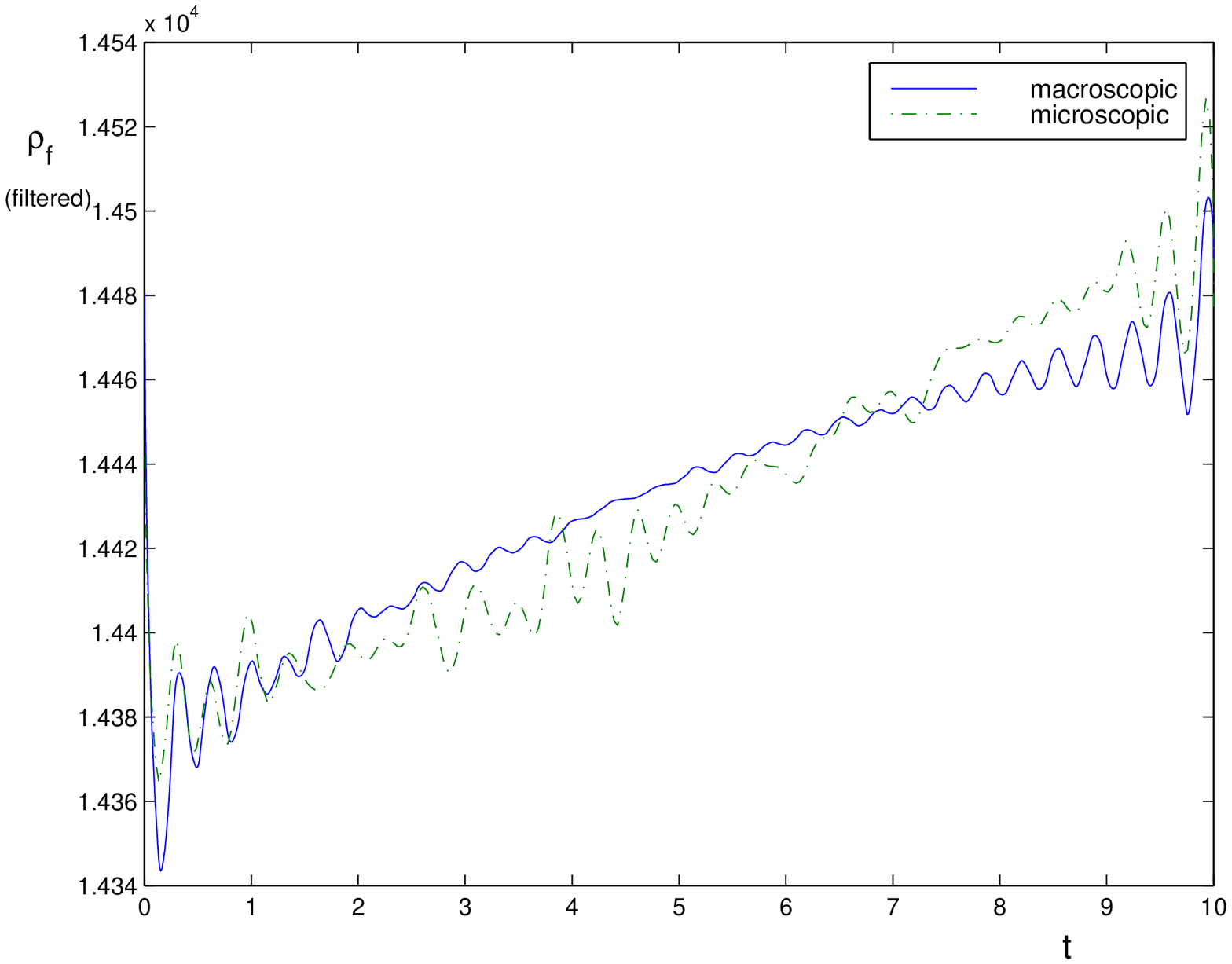}  \par{}

\caption{\label{endens2}The mean value of the energy density of the fluctuations in
the case \protect\protect\( \phi _{0}=2m.\protect \protect \) and parameters
as in fig. (\ref{fig3}). Solid line represent the macroscopic case and the
dotted line represent the microscopic theory. }
\end{figure}

In the next set of figures (fig.(\ref{fig3}) to (\ref{endens2})), we set \( f=2 \)
in order to obtain a more pronounced damping of the amplitude of the homogeneous
mode of the field. Note in figure 7 that the macroscopic model captures very
effectively the damping of the amplitude for large time. None of the parameters
of the fluid model was changed, that is the same value of \( \alpha  \) that
was used in the case \( f=1 \) was retained, namely \( \alpha =25.9 \), \( a=0 \),
\( T_{min}=10.198 \) and \( A_{0}=0.9 \) in order to show the predictive power
of the model.. Under the rather crude assumptions that lead to our model, the
result is quite satisfactory.

In figure (\ref{phidot}), we compare the time derivative of the scalar and
the macroscopic model. Finally, we depicted in figure (\ref{endens2}), the
energy density for the fluctuations in both models in the case \( f=2. \)

We conclude that the macroscopic model is able to describe the evolution of
the field and the energy density of the fluctuations with the accuracy required
for cosmological applications. 


\subsection*{Appendix A: Divergence Type Theories}

Following Geroch \cite{1}, divergence type theories are usually described in
terms of some tensorial quantities that obey conservation equations 
\begin{eqnarray}
T_{\; ;b}^{ab} & = & 0\\
N_{\; ;a}^{a} & = & 0\label{sup87} \\
A_{\quad ;a}^{abc} & = & I^{bc}
\end{eqnarray}

This is a simple and slight generalization of relativistic fluid theories proposed
initially by Liu, Muller and Ruggeri \cite{2}. In this setting, \( T^{ab} \)
is the energy-momentum tensor and \( N^{a} \) is the particle current. Their
corresponding equation simply expresses conservation of energy, momentum and
mass. The third equation will describe the dissipative part. The energy-momentum
tensor is symmetric and \( A^{abc}=A^{acb}; \) \( A_{\quad b}^{ab}=0 \) and
\( I_{\, \, ;a}^{a}=0. \) The entropy current is enlarged to read 
\begin{equation}
\label{sup88}
S^{a}=\chi ^{a}-\xi _{b}T^{ab}-\xi N^{a}-\xi _{bc}A^{abc}
\end{equation}

The \( \xi ,\xi ^{a},\xi ^{ab} \) are the dynamical degrees of freedom. The
following relations hold \cite{1}
\begin{equation}
\label{sup89}
N^{a}=\frac{\partial \chi ^{a}}{\partial \xi }
\end{equation}

\begin{equation}
\label{sup90}
T^{ab}=\frac{\partial \chi ^{a}}{\partial \xi _{b}}
\end{equation}

\begin{equation}
\label{sup91}
A^{abc}=\frac{\partial \chi ^{a}}{\partial \xi _{bc}}
\end{equation}

Symmetry of the energy-momentum tensor implies that 
\begin{equation}
\label{sup92}
\chi ^{a}=\frac{\partial \chi }{\partial \xi _{a}}
\end{equation}

That is all the fundamental tensors of the theory can be obtained from the generating
functional \( \chi . \) The entropy production is given by 
\begin{equation}
\label{sup93}
S_{\, \, ;a}^{a}=-I^{bc}\xi _{bc}
\end{equation}

Positive entropy production is ensured by demanding that \( I^{bc}=M^{(bc)(de)}\xi _{de} \),
where \( M \) is negative definite.

Ideal fluids are an important if somewhat trivial example. To obtain ideal hydrodynamics
within the DTT framework, consider a generating functional \( \chi _{p}=\chi _{p}\left( \xi ,\mu \right)  \)
where \( \mu \equiv \sqrt{-\xi _{a}\xi ^{a}}. \) It is a simple matter to obtain
\begin{equation}
\label{sup94}
\chi _{p}^{a}=-\frac{\xi ^{a}}{\mu }\frac{\partial \chi _{p}}{\partial \mu }
\end{equation}

\begin{equation}
\label{sup95}
T_{p}^{ab}=-\frac{g^{ab}}{\mu }\frac{\partial \chi _{p}}{\partial \mu }+\frac{\xi ^{a}\xi ^{b}}{\mu ^{2}}\left[ -\frac{1}{\mu }\frac{\partial \chi _{p}}{\partial \mu }+\frac{\partial ^{2}\chi _{p}}{\partial \mu ^{2}}\right] 
\end{equation}

A simple comparison with the perfect fluid form of the energy-momentum tensor
\( T^{ab}=g^{ab}p+u^{a}u^{b}\left[ p+\rho \right]  \) implies the following
identification 
\begin{equation}
\label{pressure}
p=-\frac{1}{\mu }\frac{\partial \chi _{p}}{\partial \mu }
\end{equation}

\begin{equation}
\label{sup96}
\rho =\frac{\partial ^{2}\chi _{p}}{\partial \mu ^{2}}
\end{equation}

Note that the conserved current can be quite generally written as 
\begin{equation}
\label{current}
N^{a}=\frac{\partial }{\partial \xi }\left( -\frac{\xi ^{a}}{\mu }\frac{\partial \chi }{\partial \mu }\right) =\xi ^{a}\frac{\partial p}{\partial \xi }
\end{equation}

A less trivial but important example both historically and conceptually is the
Eckart theory which can be obtained from \cite{1}
\begin{equation}
\label{sup97}
\chi _{E}=\chi _{p}+\frac{1}{2}\zeta _{ab}u^{a}u^{b}
\end{equation}

Performing a Legendre transform to the new variables \( \xi \, \xi ^{a},\xi ^{ab} \)
one obtains a system of first order differential equations of the form 
\begin{equation}
\label{sup98}
\frac{\partial ^{2}\chi ^{a}}{\partial \xi _{A}\partial \xi _{B}}\xi _{B;a}=I^{A}
\end{equation}

where \( \xi _{A} \) stand for the entire collection of variables \( \left( \xi ,\xi _{a}\, ,\xi _{ab}\right)  \)
and similarly \( I^{A}\equiv \left( 0,0,I^{ab}\right)  \) represent the dissipative
source; the index \( A \) thus covers \( 14 \) dimensions in our example.
This first order system of differential equations is symmetric since 
\begin{equation}
\label{sup99}
\frac{\partial ^{2}\chi ^{a}}{\partial \xi _{A}\partial \xi _{B}}=\frac{\partial ^{2}\chi ^{a}}{\partial \xi _{B}\partial \xi _{A}}
\end{equation}

Note that we have a system of the form 
\begin{equation}
\label{sup100}
A^{i}v_{,i}+Bv=0
\end{equation}

where \( i \) is a space-time index, the \( A^{i} \) and \( B \) are \( k\times k \)
matrices and \( v \) is a \( k \) -vector. Now this (first order) system is
hyperbolic if all its eigenvalues are real; each of these eigenvalues represent
the velocity of propagation of some small disturbance in space. These in turn
propagate along hypersurfaces called characteristics whose existence is insured
by the existence of \( k \) real eigenvalues \cite{3}\cite{4}. If the matrices
\( A^{i} \) and \( B \) are symmetric then it suffices that some combination
\( A^{i}v_{i} \) be definite (negative-definite given our choice of the signature
for the metric) to insure that all the eigenvalues are real (but some could
be degenerate). An usual case happens when this combination reduces to \( A^{0} \)
, the vector \( v \) being the time-like vector \( \left( 1,\overrightarrow{0}\right) . \)
In a relativistic theory one would expect hyperbolicity to be invariant under
(proper) Lorentz transformations; in this case we say the system is causal.
In our context, one would thus say that the system is hyperbolic if 
\begin{equation}
\label{sup101}
\frac{\partial ^{2}\chi ^{a}}{\partial \xi _{A}\partial \xi _{B}}v_{a}
\end{equation}
 is negative-definite for some temporal vector \( v_{a} \) and the theory will
be causal if this stay true for \emph{any} temporal vector \( v_{a}. \)

\subsection*{Appendix B: Computation of the expectation value of the energy-momentum tensor:}

We begin with 
\begin{eqnarray}
<T^{\mu \nu }> & = & \frac{2}{\sqrt{-g}}\frac{\delta \Gamma }{\delta g_{\mu \nu }}\\
 & = & \frac{2}{\sqrt{-g}}\frac{\delta S_{2}}{\delta g_{\mu \nu }}+\frac{2}{\sqrt{-g}}\frac{\delta \Gamma _{f}}{\delta g_{\mu \nu }}\label{sup102} 
\end{eqnarray}

using the well known identities

\begin{eqnarray}
\frac{\delta \sqrt{-g}}{\delta g_{\mu \nu }} & = & \frac{1}{2}\sqrt{-g}\: g^{\mu \nu }\\
\frac{\delta g^{\alpha \beta }}{\delta g_{\mu \nu }} & = & -\frac{1}{2}\left( g^{\alpha \mu }g^{\beta \nu }+g^{\alpha \nu }g^{\beta \mu }\right) \label{sup103} 
\end{eqnarray}

This leads to 
\begin{equation}
\label{sup104}
T_{f}^{\mu \nu }=\left[ \partial ^{\prime \mu }\partial ^{\nu }-\frac{1}{2}\eta ^{\mu \nu }\partial ^{\prime }_{\sigma }\partial ^{\sigma }\right] \left. G_{T}(x,x^{\prime })\right| _{x=x^{\prime }}-\frac{1}{2}\eta ^{\mu \nu }K\, G_{T}(x,x)
\end{equation}

with the prime denoting differentiation with respect to \( x^{\prime } \) and
where 
\begin{equation}
\label{sup105}
G_{T}(x,x^{\prime })=k_{B}T_{f}\int \frac{d^{3}k}{(2\pi )^{3}}\frac{1}{2\omega ^{2}_{k}(0)}\exp \left[ i\vec{k}\cdot (\vec{x}-\vec{x}^{\prime })\right] \left[ U_{k}(t)U^{*}_{k}(t^{\prime })+U^{*}_{k}(t)U_{k}(t^{\prime })\right] 
\end{equation}

with the limit \( \hbar \rightarrow 0 \) already taken care of. Let us consider
first the case \( \mu =\nu =0. \) We have 
\begin{eqnarray}
\partial ^{\prime }_{t}\partial _{t}-\frac{1}{2}\eta ^{00}\left( -\partial ^{\prime }_{t}\partial _{t}+\partial ^{\prime }_{i}\partial _{i}\right) \left. G_{T}(x,x^{\prime })\right| _{x=x^{\prime }} & = & \left( \frac{1}{2}\partial ^{\prime }_{t}\partial _{t}+\frac{1}{2}\partial ^{\prime }_{i}\partial _{i}\right) \left. G_{T}(x,x^{\prime })\right| _{x=x^{\prime }}\\
 & = & k_{B}T_{f}\int \frac{d^{3}k}{(2\pi )^{3}}\frac{1}{2\omega ^{2}_{k}(0)}\left\{ \left| \dot{U}_{k}(t)\right| ^{2}+|\vec{k}|^{2}\left| U_{k}(t)\right| ^{2}\right\} \label{sup106} 
\end{eqnarray}

Therefore 
\begin{equation}
\label{sup107}
T_{f}^{00}=k_{B}T\int \frac{d^{3}k}{(2\pi )^{3}}\frac{1}{2\omega ^{2}_{k}(0)}\left\{ \left| \dot{U}_{k}(t)\right| ^{2}+\left[ |\vec{k}|^{2}+K\right] \left| U_{k}(t)\right| ^{2}\right\} 
\end{equation}

In the case \( \mu =\nu =i \) we have 
\begin{eqnarray}
\partial ^{\prime }_{i}\partial _{i}-\frac{1}{2}\eta ^{jj}\left( -\partial ^{\prime }_{t}\partial _{t}+\partial ^{\prime }_{j}\partial _{j}\right) \left. G_{T}(x,x^{\prime })\right| _{x=x^{\prime }} & = & \left( \partial ^{\prime }_{i}\partial _{i}+\frac{1}{2}\partial ^{\prime }_{t}\partial _{t}-\frac{1}{2}\sum _{j=1}^{3}\partial ^{\prime }_{j}\partial _{j}\right) \left. G_{T}(x,x^{\prime })\right| _{x=x^{\prime }}\\
 & = & k_{B}T_{f}\int \frac{d^{3}k}{(2\pi )^{3}}\frac{1}{2\omega ^{2}_{k}(0)}\left\{ \left( 2k_{i}^{2}-|\vec{k}|^{2}\right) \left| U_{k}(t)\right| ^{2}+\left| \dot{U}_{k}(t)\right| ^{2}\right\} \\
 & = & k_{B}T_{f}\int \frac{d^{3}k}{(2\pi )^{3}}\frac{1}{2\omega ^{2}_{k}(0)}\left\{ \left| \dot{U}_{k}(t)\right| ^{2}-\frac{1}{3}|\vec{k}|^{2}\left| U_{k}(t)\right| ^{2}\right\} \label{sup108} 
\end{eqnarray}

Thus 
\begin{equation}
\label{sup109}
T^{ii}_{f}=k_{B}T_{f}\int \frac{d^{3}k}{(2\pi )^{3}}\frac{1}{2\omega ^{2}_{k}(0)}\left\{ \left| \dot{U}_{k}(t)\right| ^{2}-\left( K+\frac{1}{3}|\vec{k}|^{2}\right) \left| U_{k}(t)\right| ^{2}\right\} 
\end{equation}

The classical energy-momentum tensor of a scalar field is given by 
\begin{equation}
\label{sup110}
T_{c}^{\mu \nu }=\frac{2}{\sqrt{-g}}\frac{\delta S_{2}}{\delta g_{\mu \nu }}
\end{equation}

with 
\begin{equation}
\label{sup111}
S_{2}=\int \left\{ -\frac{1}{2}g_{\mu \nu }\partial ^{\mu }\phi ^{i}\partial ^{\nu }\phi ^{i}+\frac{K^{2}}{2\lambda }-\frac{1}{2}K\phi ^{i}\phi ^{i}-\frac{m^{2}}{\lambda }K\right\} \sqrt{-g}d^{4}x
\end{equation}

where we factored out a factor of \( N \). We obtained 
\begin{equation}
\label{sup112}
T_{c}^{\mu \nu }=\eta ^{\alpha \mu }\eta ^{\beta \nu }\partial _{\alpha }\phi \partial _{\beta }\phi -\frac{1}{2}\eta ^{\mu \nu }\partial _{\sigma }\phi \partial ^{\sigma }\phi +\eta ^{\mu \nu }\left\{ \frac{K^{2}}{2\lambda }-\frac{1}{2}\phi ^{2}K-\frac{m^{2}}{\lambda }K\right\} 
\end{equation}

that is 
\begin{eqnarray}
T^{00}_{c} & = & \frac{1}{2}\left( \partial _{t}\phi \right) ^{2}+\frac{1}{2}\left( \vec{\nabla }\phi \right) ^{2}+\frac{1}{2}\phi ^{2}K-\frac{K^{2}}{2\lambda }+\frac{m^{2}}{\lambda }K\\
T^{ii}_{c} & = & \frac{1}{2}\left( \partial _{t}\phi \right) ^{2}\vec{+}\left( \partial _{i}\phi \right) ^{2}-\frac{1}{2}\left( \vec{\nabla }\phi \right) ^{2}-\frac{1}{2}\phi ^{2}K+\frac{K^{2}}{2\lambda }-\frac{m^{2}}{\lambda }K\label{sup113} 
\end{eqnarray}
 which leads to 
\begin{eqnarray}
<T^{00}> & = & \frac{1}{2}p^{2}+k_{B}T\int \frac{d^{3}k}{(2\pi )^{3}}\frac{1}{2\omega ^{2}_{k}(0)}\left\{ \left| \dot{U}_{k}(t)\right| ^{2}+\left[ |\vec{k}|^{2}+K\right] \left| U_{k}(t)\right| ^{2}\right\} +\frac{1}{2}\phi ^{2}K-\frac{1}{2\lambda }\left( K-m^{2}\right) ^{2}+\frac{m^{4}}{2\lambda }\\
<T^{ii}> & = & \frac{1}{2}p^{2}+k_{B}T_{f}\int \frac{d^{3}k}{(2\pi )^{3}}\frac{1}{2\omega ^{2}_{k}(0)}\left\{ \left| \dot{U}_{k}(t)\right| ^{2}-\left( K+\frac{1}{3}|\vec{k}|^{2}\right) \left| U_{k}(t)\right| ^{2}\right\} -\frac{1}{2}\phi ^{2}K+\frac{1}{2\lambda }\left( K-m^{2}\right) ^{2}-\frac{m^{4}}{2\lambda }\label{sup114} 
\end{eqnarray}

using 
\begin{eqnarray}
\frac{K^{2}}{2\lambda }-\frac{m^{2}}{\lambda }K & = & \frac{1}{2\lambda }\left\{ K^{2}-2m^{2}K+m^{4}-m^{4}\right\} \\
 & = & \frac{1}{2\lambda }\left\{ K^{2}-2m^{2}K+m^{4}\right\} -\frac{m^{4}}{2\lambda }\\
 & = & \frac{1}{2\lambda }\left( K-m^{2}\right) ^{2}-\frac{m^{4}}{2\lambda }\label{sup115} 
\end{eqnarray}

Now 
\begin{equation}
\label{sup116}
\frac{1}{2}\phi ^{2}K+k_{B}T\int \frac{d^{3}k}{(2\pi )^{3}}\frac{K\left| U_{k}(t)\right| ^{2}}{2\omega ^{2}_{k}(0)}=K\left\{ \frac{1}{2}\phi ^{2}+k_{B}T\int \frac{d^{3}k}{(2\pi )^{3}}\frac{\left| U_{k}(t)\right| ^{2}}{2\omega ^{2}_{k}(0)}\right\} 
\end{equation}

which, using (\ref{old11}) reduces to: 
\begin{equation}
\label{sup117}
\frac{1}{2}K\phi ^{2}+k_{B}T\int \frac{d^{3}k}{(2\pi )^{3}}\frac{K\left| U_{k}(t)\right| ^{2}}{2\omega ^{2}_{k}(0)}=K\frac{\left( K-m^{2}\right) }{\lambda }
\end{equation}

Therefore 
\begin{eqnarray}
<T^{00}> & = & \frac{1}{2}p^{2}+k_{B}T\int \frac{d^{3}k}{(2\pi )^{3}}\frac{1}{2\omega ^{2}_{k}(0)}\left\{ \left| \dot{U}_{k}(t)\right| ^{2}+|\vec{k}|^{2}\left| U_{k}(t)\right| ^{2}\right\} +K\frac{\left( K-m^{2}\right) }{\lambda }-\frac{1}{2\lambda }\left( K-m^{2}\right) ^{2}+\frac{m^{4}}{2\lambda }\\
<T^{ii}> & = & \frac{1}{2}p^{2}+k_{B}T_{f}\int \frac{d^{3}k}{(2\pi )^{3}}\frac{1}{2\omega ^{2}_{k}(0)}\left\{ \left| \dot{U}_{k}(t)\right| ^{2}-\frac{1}{3}|\vec{k}|^{2}\left| U_{k}(t)\right| ^{2}\right\} -K\frac{\left( K-m^{2}\right) }{\lambda }+\frac{1}{2\lambda }\left( K-m^{2}\right) ^{2}-\frac{m^{4}}{2\lambda }\label{sup118} 
\end{eqnarray}

leading immediately to the final answer 
\begin{eqnarray}
<T^{00}> & = & \frac{1}{2}p^{2}+k_{B}T\int \frac{d^{3}k}{(2\pi )^{3}}\frac{1}{2\omega ^{2}_{k}(0)}\left\{ \left| \dot{U}_{k}(t)\right| ^{2}+|\vec{k}|^{2}\left| U_{k}(t)\right| ^{2}\right\} +\frac{1}{2\lambda }\left( K-m^{2}\right) \left( K+m^{2}\right) +\frac{m^{4}}{2\lambda }\\
<T^{ii}> & = & \frac{1}{2}p^{2}+k_{B}T_{f}\int \frac{d^{3}k}{(2\pi )^{3}}\frac{1}{2\omega ^{2}_{k}(0)}\left\{ \left| \dot{U}_{k}(t)\right| ^{2}-\frac{1}{3}|\vec{k}|^{2}\left| U_{k}(t)\right| ^{2}\right\} -\frac{1}{2\lambda }\left( K-m^{2}\right) \left( K+m^{2}\right) -\frac{m^{4}}{2\lambda }\label{sup119} 
\end{eqnarray}


\begin{thebibliography}{10}
\bibitem{preheating}There is a vast literature on preheating. Some representative works are L. Kofman,
A. Linde and A. Starobinsky, Phys. Rev. Lett \textbf{73,} 3195 (1994); Y. Shtanov,
J. Traschen and R. Brandenberger, Phys. Rev. \textbf{D51,} 5438 (1995); P. Greene,
L. Kovman, A. Linde and A. Starobinsky, Phys. Rev. \textbf{D56,} 6175 (1997);
B. Bassett, D. Kaiser and R. Maartens, Phys. Lett. \textbf{B455}, 84 (1999);
B. Bassett, C. Gordon, R. Maartens and D. Kaiser, Phys. Rev. \textbf{D61,} 061302
(2000). 
\bibitem{Bassett}S. Tsujikawa and B. Bassett, Phys. Lett. B 536, 9 (2002) 
\bibitem{D.Boyanovsky}D. Boyanovsky, H. J. de Vega, R. Holman and J. F. J. Salgado, Phys. Rev. D54,
7570 (1996); H. Fujisaki, K. Kumekawa, M. Yamaguchi and M. Yoshimura, Phys.
Rev. D53, 6805 (1996) 
\bibitem{Ramseys}S. A. Ramsey and B. L. Hu, Phys. Rev. D56, 661 (1997); S. A. Ramsey and B. L.
Hu, Phys. RevD56, 678 (1997) 
\bibitem{paper0}S. Khlebnikov and I. Tkachev, Phys. Rev. Lett \textbf{77,} 219 (1996). 
\bibitem{paper1}S. Khlebnikov, in ''Strong and Electroweak Matter '97'', e-print hep-ph/9708313v2 
\bibitem{paper2}G. Felder and I. Tkachev, e-print hep-ph/0011159 
\bibitem{paper2b}G. Felder and L. Kofman, Phys. Rev. \textbf{D63,} 103503 (2001). 
\bibitem{paper3}G. Felder, J. Garcia-Bellido, P. Greene, L. Kofman, A. Linde and I. Tkachev,
Phys. Rev. Lett \textbf{87,} 011601 (2001). 
\bibitem{Son}D. T. Son, e-print hep-ph/9601377. 
\bibitem{Finelli}F. Finelli and S. Khlebnikov, Phys. Rev. D65, 43505 (2002) J. Zibin, R. Brandenberger
and D. Scott, Phys. Rev. D63, 43511 (2001) 
\bibitem{Ashfordi}N. Afshordi and R. Brandenberger, Phys. Rev. D63, 123505 (2001) T. Tanaka and
B. Bassett, astro-ph/0302544; N. Bartolo, S. Matarrese and A. Riotto, astro-ph/0308088 
\bibitem{fof}W. Hiscock and L. Lindblom, \textit{Ann. Phys. (NY)} \textbf{151}, 466 (1983) 
\bibitem{fof2}W. Hiscock and L. Lindblom, \textit{Phys. Rev.} \textbf{D31,} 725 (1985); 
\bibitem{Kraichnan}R.Kraichnan, Dynamics of Nonlinear Stochastic Systems, J. Math.Phys. 2, 124(1961) 
\bibitem{Grana}M. Graña and E. Calzetta, Phys. Rev. D65, 63522 (2002) 
\bibitem{CHKMPA94}F. Cooper, S. Habib, Y. Kluger, E. Mottola, J. P. Paz and P. Anderson, Nonequilibrium
quantum fields in the large-N expansion, Physical Review D50, 2848 (1994) 
\bibitem{CKMP95}F. Cooper, Y. Kluger, E. Mottola and J. P. Paz, Quantum evolution of disoriented
chiral condensates, Physical Review 51, 2377 (1995) 
\bibitem{BVHS99}D. Boyanovsky, H. J. de Vega, R. Holman and J. F. J. Salgado, Nonequilibrium
Bose-Einstein condensates, dynamical scaling, and symmetric evolution in the
large \( N \) \( \Phi ^{4} \) theory, Phys. Rev. D59, 125009 (1999). 
\bibitem{LMR02}F. Lombardo, F. Mazzitelli and R. Rivers, Decoherence in Field Theory: General
couplings and slow quenches, hep-ph/0204190. 
\bibitem{HarHor81}J. B. Hartle and G. Horowitz, Ground-state expectation value of the metic in
the 1/N or semiclassical approximation to quantum gravity, Phys. Rev. D24, 257
(1981). 
\bibitem{MP89}F. D. Mazzitelli and J. P. Paz, Gaussian and 1/N approximation in semiclassical
cosmology, Phys. Rev. D39, 2234 (1989) 
\bibitem{RY00}A. Ryzhov, L. Yaffe, Large N quantum time evolution beyond leading order, Phys.
Rev. D62, 125003 (2000) 
\bibitem{AABBS02}G. Aarts, D. Ahrensmeier, R. Baier, J. Berges and J. Serreau, Far-from-equilibrium
dynamics with broken symmetries from the \( 2PI-1/N \) expansion, hep-ph/0201308. 
\bibitem{BS03}J. Berges and J. Serreau, Progress in nonequilibrium quantum field theory, hep-ph/0302210;
D. Boyanovsky, C. Destri and H. J. de Vega, hep-ph/0306124; S. Juchem, W. Cassing
and C. Greiner, hep-ph/0307353 
\bibitem{Hu}E. Calzetta and B. L. Hu, hep-ph/0305326 (to appear in Phys. Rev. D) 
\bibitem{BS02}J. Berges and J. Serreau, hep-ph/0208070 (to appear in Phys. Rev. Lett.) 
\bibitem{Baacke}J. Baacke, D. Boyanovsky and H. J. de Vega, Phys. Rev. D63, 45023 (2001) 
\bibitem{ours2}E. Calzetta and M. Thibeault, Phys. Rev. D63, 103507 (2001) E. Calzetta and
M. Thibeault, Int. J. Theor. Phys 41, 2179 (2002) 
\bibitem{Geroch}R. Geroch and L. Lindblom, Phys.Rev D \textbf{41} 6, 1855 (1990) 
\bibitem{Tkachev00}G.Felder and I. I. Tkachev, hep-ph/0011159 
\bibitem{Starobinsky95}D. Polarski and A. Starobinsky, gr-qc/9504030 
\bibitem{Tkachev96}S. Y. Khlebnikov and I. I. Tkachev, hep-ph/9603378 
\bibitem{Huang}K. Huang, \emph{Statistical Mechanics} (John Wiley, New York (1987)) 
\bibitem{israel88}W. Israel, \textit{Covariant fluid mechanics and thermodynamics: an introduction},
in A. Anile and Y. Choquet - Bruhat (eds.) \textit{Relativistic fluid dynamics}
(Springer, New York, 1988). 
\bibitem{Calzetta}E. Calzetta, Clas.Quantum Grav. \textbf{15,} 653 (1998) 
\bibitem{1}R. Geroch, and L. Lindblom (1990).Physical Review D \textbf{14}, 1855. 
\bibitem{2}I. Liu, I. Muller and T. Ruggeri,Ann.Phys.(N.Y.)\textbf{169},191 
\bibitem{3}R. Courant and D. Hilbert, \emph{Methods of Mathematical Physics II}, Partial
Differentials Equations (Interscience, New York, 1962) 
\bibitem{4}J. Stewart, \emph{Advanced General Relativity}, Cambridge Monographs on Mathematical
Physics (Cambridge University Press, Cambridge, England, 1993) 
\bibitem{Aarts}G. Aarts and J. Smit, Phys. Lett. B393 (1997),{[}hep-ph/9610415{]}
\bibitem{numerical}W. H. Press, S. A. Teukolsky, W. T. Vettering and B. P. Flannery, \emph{Numerical
Recipes in Fortran}, 2d edition, (Cambridge University Press, Cambrige, England,
1992) 
\end{thebibliography}
\end{document}